# Theory of Crystallization under Equilibrium Polymerization in a Solution and the Investigation of its Melting Properties

**Sagar S. Rane[*] and P. D. Gujrati[*†]**

**Department of Polymer Science,[*] Department of Physics,[†]**

**The University of Akron, Akron, OH 44325**

## Abstract

We generalize a recently investigated lattice model of semiflexible polymers formed under equilibrium polymerization in a solution and conduct a comprehensive investigation of its melting properties. The model is characterized by six energies, three of which are for the interaction between the middle-group, the end-group and the solvent, and the remaining three represent energies for a gauche bond, a hairpin turn and a pair of neighboring parallel bonds. A polymer has two end-groups and at least one chemical bond. Two activities control the end-group and the middle-group densities, respectively, and give rise to polydisperse chains whose number is not fixed. We study the melting properties with various model parameters under conditions of fixed pressure, and compare our results with experimental data on fixed length and polydisperse polymers, whenever available. We investigate the effect of monomer interactions, nature of end-groups, chain rigidity, solvent quality, degree of polymerization, etc on the melting properties such as the melting temperature, latent heat, and energy and entropy of fusion. Our theory is thermodynamically consistent in the entire parameter space and improves upon the classical theories; hence our results should prove useful.

# I. Introduction

The technological importance and the complexity of modeling polymer crystallization have drawn the interest of several workers,[1-25] where the interest has been to study equilibrium melting. Most workers have studied the original Flory model of melting[2] because of its simplicity. It is fair to say that there is no consensus as to the nature of the transition in the model;[5-13] see, for example, Ref. 16. It has also become clear[19-21] that the semiflexibility alone, which is all that is present in the Flory model,[2] is *not* sufficient to give a first-order melting. Thus, the understanding of equilibrium melting is not complete at this moment. Experimental investigation of melting of macromolecules is complicated because the free energy barrier between the kinetically selected crystalline state and the true equilibrium state is so large that the kinetically selected metastable crystal state persists over time scales much longer than the experimental time scale.[17] It is also clear that non-equilibrium features *cannot* be understood well without a comprehensive understanding of equilibrium melting, which is far from complete. Therefore, this work is limited to investigating equilibrium melting only.

In this paper, we study the crystallization of linear chains produced under the condition of equilibrium polymerization in a solution. Here, polymers can break *anywhere* along their backbone sequence including the ends, and recombine at their ends reversibly in such a way that they maintain a particular *equilibrium distribution* of their lengths[26] Even the number of chains is not fixed. Thus, one deals with a specific *polydispersity* in the chain length distribution, and number distribution. This should be contrasted with the conditions under which living polymers are formed, where polymers grow *stepwise* only at their ends under conditions that need not reach equilibrium.[27] The number of polymers remains *fixed* during living polymerization, whereas it *is not* fixed during equilibrium polymerization, though its average is. Thus, living polymerization is in equilibrium with respect to chemical bonds, but not with respect to the number of polymers. Despite the differences, many authors treat living polymerization as equilibrium polymerization[14,27,28] as defined above. This issue has been discussed carefully in Ref. 29, to which we direct the reader. We will, henceforth, not make this association in this work.

Equilibrium polymerization of *athermal* linear polymers has been widely studied,[30,31] due to its deep connection with the $n$=0 limit of an $n$-vector isotropic magnetic model. We will closely follow the connection developed in Ref. 31, where the weight of a solvent molecule is independent of the magnetic field $H$ in the magnetic system. The magnetic mapping is also extended to the polymerization model of randomly branched athermal polymers of "even" functionality.[32] Following this connection, a through understanding of equilibrium polymerization of linear chains and branched polymers in athermal solutions has developed over the years, either with the use of the $n$=0 limit[30-33] or without it.[28,34-43] The latter approach also allows consideration of polymer interactions. Equilibrium polymerization of (interacting) branched polymers also describe thermoreversible gelation, and have been studied extensively by many authors.[36,38-46] However, all these studies are related to completely flexible polymers.

The study of crystallization, which requires semi-flexibility, in linear chains under equilibrium polymerization has so far received little attention, though some important attempts have been made.[13-15,21,23-25] Most of the work is either numerical, or phenomenological; in the latter case, the work is based on the random mixing approximation used in the Flory theory.[2] However, not much information is available on how the melting properties of such polymers relate to the molecular parameters. Our goal in this paper is to fill in this gap by developing a theory for crystallization under equilibrium polymerization in a solution that goes beyond the random mixing approximation that is used in the classical Flory theory of melting[2] and conducting a comprehensive study of the melting properties. The model we use here is an extension of the model proposed recently;[19,20] the extended model has already been used to study the role of free volume on the ideal glass transition in Ref. 21.

The first derivation of lattice statistics in a model of semiflexible polymers with the mean field approximation was given by Flory,[1,2] where monodisperse polymers were considered. Flory has suggested that melting is mainly dictated by the excluded volume interaction, and chain rigidity and intermolecular interactions can be neglected. This suggestion has been criticized[4] from the results of the exactly solvable models such as the KDP, F and dimer models by mapping them onto polymer models; see also Refs. 8, and 16. Huggins obtained a better estimate for the probability of monomer insertion, which was shown to give more reliable[4,6] results for the melting temperature than the original Flory calculation. In a significant development, the results of Flory's theory were shown to be wrong, not just quantitatively but more important qualitatively by Gujrati and Goldstein.[5-7] We refer the reader to Refs. 6, 8, and 20 for further details. In particular, it was shown that no low temperature inactive phase with perfect order can exist at nonzero temperatures ($T$>0) in the Flory model because of the presence of the Gujrati-Goldstein excitations[5-8] involving hairpin turns, and the density of gauche bonds goes to zero only at $T$=0. In an exact calculation on a Husimi cactus, which looks identical to a square lattice locally, it has been found that the melting transition in the Flory model is a tricritical point.[19,20] By introducing inter- and intra-molecular interactions, we can obtain a first order melting.[19-21] suggesting strongly that the inter- and intra-molecular interactions may be very important in determining the order of the transition and, hence, must be accounted for.[4,8]

The equilibrium polymerization is best described by a partition function. Such a partition function for the general polydisperse model of linear chains is presented in Ref. 20, (see their Eq. 7) where we allow for semiflexibility, inter- and intra-molecular interactions, polydispersity (controlled by the end-group activity) and solvent or free volume. No distinction is made between the functional end groups and the middle-groups of polymers. However, depending upon the functional groups used in initiation and termination, the end-groups may be similar or very different from the middle-groups. The importance of accounting for end-group effects was recently demonstrated by us.[47] Therefore, we will extend the above model[20] to incorporate the difference between the end-groups and the middle-groups. The model is fully defined in the next section and has been used in Ref. 21 in a different context to investigate metastability and the ideal glass transition. Here, our interest is to investigate melting properties. We consider a square lattice so as to mimic a tetrahedral lattice needed to describe polymer crystals.

The model *cannot* be solved exactly. Hence, we make one single approximation, after which we solve the model exactly: We replace the original square lattice with a Husimi cactus;[19] which locally looks similar to the square lattice as both contain a simple square as the basic unit; see Fig. 1. The squares are connected in a tree-like fashion on the cactus, which makes the cactus different from the square lattice. The tree nature of the cactus allows us to solve the model *exactly*. The resulting solution is taken as an approximate theory for the original square lattice. The use of the Husimi cactus allows us to incorporate Gujrati-Goldstein excitations[5-8,19-21] that are responsible for destroying the completely ordered crystal phase in the Flory model. Thus, our theory is an improvement over the Flory theory. The Husimi cactus also takes into account more correlations than the Bethe lattice[48] (another recursive lattice) due to the finite loop size. The Huggins approximation has been shown[4] to be exact on the Bethe lattice for a fully packed system. The Flory approximation for the probability of monomer insertion is also exact in the limit of the coordination number $q \to \infty$.[4] The Husimi cactus turns into the Bethe lattice as the loop size goes to infinity.[49] Thus, we also expect our theory to give better results than the Huggins approximation. The method of solution of our model is to use recursion relation technique proposed elsewhere.[48] We identify the crystalline state at absolute zero by a novel two cycle fixed point scheme, which has been presented earlier.[19-21] The perfectly ordered crystalline state at absolute zero in the present theoretical treatment is of infinite size and consists of parallel arrangement of chains. At this time, we do not consider finite size crystallites or any particular type of unit cell or structure for the polymer; this is similar to the approach used by Flory.[2]

In the next two sections, we introduce the model and present the theory we shall use for our study. In section IV, we present our results for the melting properties and compare them with other literature data. The last section contains the conclusions and a brief summary of our results. In the Appendix, we present the recursion relations and other equations used in the calculations.

## II. Model

We consider, for simplicity, a system consisting of solvent molecules and polydisperse linear polymers on a square lattice of $N$ sites. The end-groups are treated as a different species from the middle-groups. We will use monomers to collectively denote the middle- and the end-groups. Each monomer or solvent molecule occupies a site of the lattice. All lattice sites are occupied. Thus, we consider an incompressible polymer solution. It is also possible to think of the solvent molecules as representing voids. In that case, we have a compressible pure system of polydisperse polymers. The excluded volume effects are accounted for by imposing the requirement that only one monomer or solvent molecule can occupy a site of the lattice. For simplicity, we only allow non-bonded interactions between nearest-neighbor sites (the exchange energy) occupied by unlike species and between next-nearest sites (the configurational energy) occupied by monomers (next-nearest neighbor monomers). Consider the possible states of the four sites that belong to a square cell of the lattice. There are five distinct states $\sigma$ that we need to consider, as shown in Fig. 2:

(a) $\sigma$=0: There is no polymer bond inside the cell.

(b) $\sigma$=1: Two neighboring sites are occupied by one polymer bond.

(c) $\sigma$=g: Two polymer bonds requiring three sites are connected to each other, making a bend.

(d) $\sigma$=p: Two polymer bonds, each requiring two sites, are parallel to each other.

(e) $\sigma$=h: Three polymer bonds are connected together to make a hairpin turn.

For a square lattice, which has a coordination number $q$=4, there are $2N$ lattice bonds, provided we neglect the surface corrections. Let $N_\sigma$ denote the number of square cells in state σ. The total number of cells is $N$; we neglect surface effects. We use "v" to denote the solvent (or voids) henceforth. Let $N_{\mathrm{M}}, N_{\mathrm{E}}$, $N_{\mathrm{v}}$, and $N_{\mathrm{m}} \equiv N - N_{\mathrm{v}}$ denote the number of sites that are occupied by the middle-groups, end-groups, solvent species and all the monomers, respectively. Then we have

$$N_{\mathrm{M}} + N_{\mathrm{E}} + N_{\mathrm{v}} \equiv N\,. \tag{2.1}$$

In the thermodynamic limit $N \to \infty$, $N_{\mathrm{M}}, N_{\mathrm{E}}$ and $N_{\mathrm{v}}$ diverge such that the corresponding densities $\phi_{\mathrm{M}} = N_{\mathrm{M}}/N, \phi_{\mathrm{E}} = N_{\mathrm{E}}/N$, and $\phi_{\mathrm{v}} = N_{\mathrm{v}}/N$ ($\phi_{\mathrm{m}} \equiv 1 - \phi_{\mathrm{v}}$) are kept fixed. Thus, Eq. (2.1) can be written in terms of densities as follows,

$$\phi_{\mathrm{M}} + \phi_{\mathrm{E}} + \phi_{\mathrm{v}} \equiv 1\,. \tag{2.2}$$

Let $B$ and $p$ denote the total number of chemical bonds in all polymers and the total number of polymer chains, respectively. If $N_{\mathrm{vv}}, N_{\mathrm{EE}}, N_{\mathrm{MM}}, N_{\mathrm{Mv}}, N_{\mathrm{Ev}}$ and $N_{\mathrm{ME}}$ denote the number of unbonded contacts between solvent-solvent, end-group/end-group, middle-group/middle-group, middle-group/solvent, end-group/solvent and middle-group/end-group pairs respectively, then it is easy to verify the following topological identities:

$$\begin{aligned}
&N = N_0 + N_1 + N_{\mathrm{g}} + N_{\mathrm{p}} + N_{\mathrm{h}}\,,\\
&2B = N_1 + 2(N_{\mathrm{g}} + N_{\mathrm{p}}) + 3N_{\mathrm{h}}\,,\\
&4N_{\mathrm{v}} = 2N_{\mathrm{vv}} + N_{\mathrm{Mv}} + N_{\mathrm{Ev}}\,,\\
&4N_{\mathrm{M}} = 2(B - p) + 2N_{\mathrm{MM}} + N_{\mathrm{Mv}} + N_{\mathrm{ME}}\,,\\
&4N_{\mathrm{E}} = 2p + 2N_{\mathrm{EE}} + N_{\mathrm{ME}} + N_{\mathrm{Ev}}\,,\\
&2p = N_{\mathrm{E}}\,,\\
&2(N_{\mathrm{MM}} + N_{\mathrm{vv}} + N_{\mathrm{EE}} + N_{\mathrm{Mv}} + N_{\mathrm{Ev}} + N_{\mathrm{ME}}) = 4N_0 + 3N_1 + 2(N_{\mathrm{g}} + N_{\mathrm{p}}) + N_{\mathrm{h}}\,.
\end{aligned} \tag{2.3}$$

From Eqs. (2.1) and (2.3), it is easily seen that the following identity,

$$2N = B + N_{\mathrm{Mv}} + N_{\mathrm{Ev}} + N_{\mathrm{ME}} + N_{\mathrm{MM}} + N_{\mathrm{EE}} + N_{\mathrm{vv}}, \qquad (2.4)$$

is satisfied as expected. We have sixteen quantities and there are eight constraints; see Eqs. (2.1), and (2.3). We choose the following eight quantities as independent in order to describe every state in our model: $N_{\mathrm{E}}, N_{\mathrm{M}}, N_{\mathrm{g}}, N_{\mathrm{p}}, N_{\mathrm{h}}, N_{\mathrm{Mv}}, N_{\mathrm{Ev}}$ and $N_{\mathrm{ME}}$. Corresponding to each of these independent quantities there exists a chemical potential or an interaction energy, which controls their number. We introduce a three-site bending penalty $\varepsilon > 0$ for each of the two possible gauche (g) bonds at each site of the lattice. There is no penalty for a trans bond. There is a four-site interaction of energy $\varepsilon' > 0$ for each pair of neighboring parallel bonds. For the hairpin turn, we have a third energy of interaction $\varepsilon'' > 0$ ($\varepsilon''$ is the energy of a hairpin turn over and above $2\varepsilon + \varepsilon'$). For each of the unbonded contacts $N_{ij}$, $i \neq j$=M,E,v, there exists an exchange energy of interaction $\varepsilon_{ij}$. The total energy of the system is given by,

$$E = \varepsilon N_{\mathrm{g}} + \varepsilon' N_{\mathrm{p}} + \varepsilon'' N_{\mathrm{h}} + \sum_{i \neq j = \mathrm{M,E,v}} \varepsilon_{ij} N_{ij}$$

$$= \varepsilon (N_{\mathrm{g}} + a N_{\mathrm{p}} + b N_{\mathrm{h}} + c_{\mathrm{Mv}} N_{\mathrm{Mv}} + c_{\mathrm{ME}} N_{\mathrm{ME}} + c_{\mathrm{Ev}} N_{\mathrm{Ev}}), \qquad (2.5)$$

where $a = \varepsilon'/\varepsilon$, $b = \varepsilon''/\varepsilon$, $c_{\mathrm{Mv}} = \varepsilon_{\mathrm{Mv}}/\varepsilon$, $c_{\mathrm{ME}} = \varepsilon_{\mathrm{ME}}/\varepsilon$ and $c_{\mathrm{Ev}} = \varepsilon_{\mathrm{Ev}}/\varepsilon$. It should be noted that the exchange energies could be expressed in terms of bare van der Waals energies $e_{ij}$ as follows:

$$\varepsilon_{ij} \equiv e_{ij} - (e_{ii} + e_{jj})/2, \quad i, j = \mathrm{M,E,v}. \qquad (2.6)$$

We also introduce Boltzmann weights or activities, which control the density of the above states, as follows: $w = \exp(-\beta\varepsilon)$, $w' = w^a$, $w'' = w^b$, and $w_{ij} = w^{c_{ij}}$, $i \neq j$=M,E,v, for the gauche bonds, parallel bonds, hairpin turn and middle-group/solvent, middle-group/end-group and end-group/solvent pairs, respectively. Here, $\beta = 1/\widetilde{T}$ is the inverse temperature, and $\widetilde{T}$ is measured in the units of the Boltzmann constant. We introduce a dimensionless temperature $T = \widetilde{T}/\varepsilon$, which will be used throughout the paper to measure the temperature. The density of end-groups and the middle-groups and their fluctuations are controlled by the activities $H = w^{-\mu_{\mathrm{E}}}$ and $\eta_{\mathrm{M}} = w^{-\mu_{\mathrm{M}}}$ respectively, where $\mu_{\mathrm{E}}$ and $\mu_{\mathrm{M}}$ play the role of the respective reduced chemical potentials, normalized by $\varepsilon$. The presence of the activity $H$ ensures that the polymer number is *not fixed*. The partition function for fixed $N$ is given by,

$$Z = \sum \eta_{\mathrm{M}}^{\mathrm{N_M}} H^{2p} w^{N_{\mathrm{g}}} w'^{N_{\mathrm{p}}} w''^{N_{\mathrm{h}}} \prod_{ij} w_{ij}^{N_{ij}}, \qquad (2.7)$$

where the sum is over all distinct states as characterized by the eight independent quantities. Each polymer has two end-groups. Here, $p$ represents the number of polymers and $B$ the total number of bonds in them in a given configuration. We do not allow middle-group or end-group monomers to exist by themselves. Thus, we are considering the properties of the polymers at 100% *conversion*. The smallest polymer has no middle-group. In other words, each polymer has at least one chemical bond.

It should be pointed out that neither the number of bonds in each polymer nor their number is fixed in the average state. Thus, polymers appearing in Eq. (2.7) are polydisperse. The free energy per site $z_0 = (1/N)\ln Z$ in the limit $N \to \infty$ is the adimensional osmotic pressure $Pv_0/\widetilde{T}$, across a membrane through which the solvent can pass through without any penalty, as shown elsewhere.[50] Here, $P$ is the conventional osmotic pressure, and $v_0$ the lattice cell volume. However, the reduced and adimensional osmotic pressure $P_{\rm r} \equiv Pv_0/\varepsilon$ is a convenient quantity and plays a useful role in our investigation. If the solvent is replace by voids, then the osmotic pressure becomes the pressure. As $\eta_{\rm M} \to \infty$, and $H \to 0^+$ or *p*=1, the model reduces to the Hamilton walk limit problem studied earlier.[19,20]

## III. Husimi Lattice Theory

As the model *cannot* be solved exactly on a square lattice, we resort to some approximate calculation. We make one approximation: we replace the square lattice by a Husimi cactus; see Fig. 1. We then solve the model exactly. The exact solution becomes an approximate theory on the square lattice.

The cactus is an infinitely large tree, obtained by joining two squares at each corner recursively. It is divided into generations, such that the generation number *m* at each site increases as we move away from the origin of the cactus, denoted by *m*=0. Each square has four sites: one at the base, which is close to the origin, one at the peak, and two intermediate sites. We index the base site by *m*, the two intermediate sites with (*m*+1), and the peak site with (*m*+2). In our model, there are seven distinct possible states at each site, as described below. Each site is shared by two squares $\Sigma$ and $\Sigma'$ with $\Sigma'$ closer to the origin; see Ref. 19. The site under consideration is the base site of the square $\Sigma$, and is marked by the filled dot close to the site, as shown in Fig. 3. Consider some $\Sigma$ and its base site near the filled dot. To define six of the seven states, we look from inside $\Sigma$ across its base site into the opposite square $\Sigma'$. The six states *L*, *R*, *O*, *I*, $E_{\rm b}$ and $E_{\rm a}$ correspond to a left turn, a right turn, an outside turn in $\Sigma'$, an internal turn in $\Sigma$ and an end-group at the site connected to a polymer coming from below or above that site respectively. The last state is the one where the site is occupied by a solvent molecule, such a state is denoted by *v*.

As discussed in Ref. 19, we only consider $a$<1 here, for which the ground state at $T$=0 is the one in which all bonds are parallel ( $N_{\rm p}$=$N$) with no bends ( $N_{\rm g}$=0), see 1 and 2 in Fig.3. This is a completely ordered crystalline state. [For $a$>1, the ground state is a step-like walk, see 3 and 4 in Fig. 3, with maximum number of bends ( $N_{\rm g}$=$N$); but no neighboring parallel bonds ( $N_{\rm p}$=0).[19,20] This state does not correspond to a crystal.] In the incompressible model with the Hamilton walk limit,[19] the appropriate range for $a$ was identified as between 0 and 1 by considering the ground state and the requirement for a first order melting. In the present model too, the range for the parameters will be restricted by similar requirements.

We define partial partition functions (PPF's) $Z_m(\alpha)$ at the *m*-th generation site given that the state at the site is $\alpha$=*L*, *R*, *I*, *O*, *v*, $E_{\rm b}$, $E_{\rm a}$; $Z_m(\alpha)$ is the contribution from the part of the cactus above the *m*-th site and is expressed in the form of recursion relations (RR's) in terms of the partial partition functions at higher sites with indices (*m*+1), and (*m*+2) within the square. Thus, the RR's for the PPF's can be symbolically written as $Z_m(\alpha) = J_\alpha[\{Z_{m+1}(\alpha')\},\{Z_{m+2}(\alpha'')\}]$, where $J_\alpha$ is a cubic polynomial in its arguments, and quadratic polynomial in $Z_{m+1}(\alpha')$. We introduce the seven ratios

$$x_m(\alpha) = Z_m(\alpha)/[Z_m(L)+Z_m(R)]$$

for $\alpha = L, R, I, O, v,$ $E_{\rm b}$ and $E_{\rm a}$, and

$$x_m(R) = 1 - x_m(L).$$

The RR's for $Z_m(\alpha)$ yield the RR's for $x_m(\alpha)$. On an infinite cactus, the RR's for $x_m(\alpha)$ will approach a fix-point (FP) solution near the origin. The FP solution describes the thermodynamics of the homogeneous bulk system. In case there are several FP solutions, then the most stable solution will determine the homogeneous bulk behavior in the model. As discussed earlier,[19] we will consider two different schemes for the FP to identify the amorphous and crystalline phases.

**Construction of Recursion Relations**

As an example, we show here how the PPF for state *I* (see section A of the Appendix) is obtained. We need to consider all possible distinct configurations of the square, with level *m* (the bottom vertex) in state *I*, as shown in Fig. 4. The statistical weight of each configuration of the square is given in section C of the Appendix, and $Z_m(I)$ is the obtained by summing these weights. Let us show how the statistical weight of the configuration shown in Fig. 4(i) is obtained. The two intermediate (*m*+1) levels are in state *R* and *O*, the peak (*m*+2) level is in state $E_{\mathrm{b}}$. Thus, the contribution from these states in Fig. 4(i) is $Z_{m+1}(R)Z_{m+1}(O)Z_{m+2}(E_{\mathrm{b}})$. In addition, we have to account for the two gauche bonds, a pair of parallel bonds, a hairpin turn, the interaction between the end-group and the middle-group within the square, and the presence of the middle-group at level *m*. The additional multiplicative contribution is $\eta_{\mathrm{M}} w^2 w' w'' w_{\mathrm{ME}}$. We do not have to include an activity $\eta_{\mathrm{M}}$ for the middle-groups at the (*m*+1) levels, or an activity *H* for the end-group at the (*m*+2) level, because those activities will be included in the corresponding PPF at levels (*m*+1) or (*m*+2). Therefore, the total statistical weight from the configuration in Fig. 4(i) is, $\eta_{\mathrm{M}} w^2 w' w'' w_{\mathrm{ME}} Z_{m+1}(R)Z_{m+1}(O)Z_{m+2}(E_{\mathrm{b}})$, as shown in section C of the Appendix. The statistical weights of other configurations in Fig. 4 are obtained in a similar fashion. The recursion relations for states $\alpha$=$O$, $v$, $L$, $E_\mathrm{a}$ and $E_\mathrm{b}$ are obtained similarly by considering all the distinct configurations of the square with level $m$ in state $\alpha$. We only show the configurations for state $I$ as the numbers of distinct configurations for other states are very large.

We present the recursion relations, the expression for $z_0$, calculation of various densities and the average degree of polymerization ($M$) in section A of the Appendix.

**Amorphous phase: 1-cycle FP scheme**

For the amorphous phase, the FP solution is very simple. As we approach the origin, $x_m(\alpha) \to x_\alpha$, regardless of the index $m$. We set $x_\alpha$ = $l$, $i$, $o$, $s$, $e_\mathrm{b}$ and $e_\mathrm{a}$, for $\alpha$ = $L$, $I$, $O$, $v$, $E_\mathrm{b}$, $E_\mathrm{a}$, respectively, and $x_\mathrm{R} \equiv r = 1 - l$. In the amorphous phase, due to symmetry, $Z_m(L) = Z_m(R)$, we always have $l$=$r$=½.

**Crystal phase: 2-cycle FP scheme**

The role of the 2-cycle solution has been explained earlier.[19,20] In this cycle pattern, the lattice has a sublattice structure: the sites along any chosen direction alternate between two type A and B. Thus, $l$ or $r$ on each sublattice A or B will no longer be ½. However, if $l$>½ on sublattice A (or B), then $r$>½ on sublattice B (or A), and the pattern repeats itself. The relation $l$+$r$=1 is still valid on each sublattice. At $T$=0, the crystal (CR) has $l_\mathrm{A}$=1, $l_\mathrm{B}$=0 or vice versa. The sublattice structure is not a property of the pure solvent phase at absolute zero, which coexists with the pure CR. However, we are not interested in the pure solvent phase in this work. The phase separation will always be present in real systems in which there is always a repulsive monomer-solvent interaction ($c_{\mathrm{Mv}} > 0$ and $c_{\mathrm{Ev}} > 0$). If the solvent is athermal, there is no phase separation, even at $T$=0. We do not consider this case, as it is not realistic. The free energy per site is calculated by considering the possible states at the origin using the Gujrati trick.[48] The methodology of this calculation is presented in Ref. 20, and will not be reported here.

## IV. Results and Discussion

Before we present our results, we briefly discuss the appropriate choice of parameters. As said earlier, we take 0<$a$<1. The energy barrier for the trans↔gauche rotation in polyethylene has been estimated[51] to be between 2.5-6.4 KJ/mole. The barrier to rotation of course depends on the nature of bonds, side groups, etc. that are specific to each species; however, we will use the energy barrier for polyethylene here taking it as a representative value. If we consider the solvent to represent the free volume, the exchange energy $\varepsilon_{\mathrm{Mv}} = -e_{\mathrm{MM}}/2$, where $e_{\mathrm{MM}}$ is the bare van der Waals energy of interaction between the middle groups. A good estimate for $-e_{\mathrm{MM}}$ appears to be in the range of 0.1-1.8 Kcal/mole[52] (assuming that the coordination number in the bulk is ≅ 10). Therefore, rough estimates indicate that the appropriate values for $c_{\mathrm{Mv}}$ are between 0.03 and 1.5. However, if the solvent represents a material species and not the free volume, $\varepsilon_{\mathrm{Mv}}$ will be much smaller and the higher limit on the choice of $c_{\mathrm{Mv}}$ will be much smaller; it may be close to zero. Similar arguments apply to the choice of $c_{\mathrm{Ev}}$. The ratio $c_{\mathrm{ME}}$, however, will usually be small as $\varepsilon_{\mathrm{ME}}$ is the exchange energy between two material species and will usually be small. In most of the calculations, we fix the reduced (osmotic) pressure $P_{\mathrm{r}} \cong 1$, as we get reasonable values for the solvent/free volume density in the liquid and the crystal for this pressure. The cell volume $v_0$ is often used as a fitting parameter,[53] specific to a given system; therefore, extracting the pressure $P$ from the value of $P_r$ is system dependent. In the figures, we do not show the dimension, because all those quantities are dimensionless. Both the energy ($E$) and the latent heat per monomer ($L_m$) are normalized by $\varepsilon$.

We allow for variable concentration of end-groups through its activity $H$, which is what is expected under condition of equilibrium polymerization. The absence of free monomers (100% conversion) should not be a concern, since our goal is not to study the polymerization process itself, but to study the melting properties of the final (polydisperse) polymer system with model parameters. The polydispersity gives rise to an average degree of polymerization (DP) that can be controlled by varying various parameters in the theory. For example, letting $H \rightarrow 0$, we can make the average DP→∞. We will use $M$ to denote the average degree of polymerization in the following.

**1. Middle Group Interaction:** We first study how the properties at the melting point change with the middle-group/solvent interaction ($c_{Mv}$). This interaction is the most important one as most of the polymer chain is composed of middle-groups. This interaction also defines the solvent quality. As $c_{Mv}$ increases, the solvent gradually becomes a more poor solvent. Thus our discussion on the effect of $c_{Mv}$ is also useful to provide information on the effect of solvent quality on the melting properties. In Fig. 5(a) we fix the pressure $P_r = 1$ and show how the melting temperature ($T_M$) and the latent heat per monomer ($L_m$) changes with $c_{Mv}$. We find that with increasing $c_{Mv}$, $T_M$ increases monotonically. The direct relation between the strength of monomer interactions and the melting temperature in the case of fixed length polymers has been established from experimental data for a number of systems by Bunn.[52] Thus, our result for the melting temperature is consistent with experimental data (of fixed length polymers), and provides a first principle basis for such a relation. If we consider the gauche bond energy $\varepsilon$ = 4 KJ/mole for polyethylene, we get a melting temperature of ≅ 126°C [see Fig. 5(a)] near $c_{\mathrm{Mv}} \cong 0.2$. Thus, our model makes reasonable predictions.

We have found that as $c_{\mathrm{Mv}}$ increases, the melting gradually becomes continuous, and then remains continuous. This can also be inferred from Fig. 5(a-c). In our model, CR only has liquid crystalline ordering, and does not contain any point-group symmetry of a conventional crystal. Hence, the melting in our model can be continuous. In the following, therefore, we will only consider cases where melting is first order. It is obvious that accounting for intermolecular interactions with the correct values may be important to yield a first order melting. This is consistent with our earlier observation.[19,20] A similar result was also obtained by Nagle[4] with the dimer model for polyethylene, where the melting is continuous for very strong intermolecular attractive interactions and becomes first order when the interactions are reduced to the physical range. All these observations are inconsistent with the one drawn in Ref. 10(a) and do not support Flory's hypothesis[2] that intermolecular interactions do not play an important role in melting. Bunn[52] has also argued that the energy barrier to rotation is of the same order of magnitude as the intermolecular interactions and hence both intermolecular interactions and molecular flexibility are equally important in determining the melting point and other properties.

In Fig. 5(b), we show how $M$ and $\phi_v$ in the crystal (CR) and the equilibrium liquid (EL) at the melting point change with $c_{Mv}$. The equilibrium liquid represents the liquid phase above the melting transition, and not its analytical continuation below it to represent the supercooled liquid. We notice that the degree of polymerization of CR ($M_C$) is always larger than the degree of polymerization of EL ($M_L$). Such a discontinuity in the degree of polymerization has been observed in theoretical results,[23-25] and Monte Carlo simulations[13,15] of equilibrium polymerization; although the latter calculations have been performed at fixed monomer chemical potential rather than fixed pressure. The CR free volume $\phi_{vC}$ is smaller than the EL free volume $\phi_{vL}$. With increasing $c_{Mv}$, we find that both $M_C$ and $M_L$ decrease, with the drop in $M_C$ being much larger. At $c_{Mv}$=0.4, the difference between $M_C$ and $M_L$, and between $\phi_{vC}$ and $\phi_{vL}$, is about 100, and 0.01 respectively. The CR free volume increases while $\phi_{vL}$ decreases with increasing $c_{Mv}$, with the change in $\phi_{vL}$ being much larger.

In Fig. 5(c), we show the entropy and energy per monomer in CR ($S_C$ and $E_C$) and EL ($S_L$ and $E_L$) at the melting point for various $c_{Mv}$. We find that with increasing $c_{Mv}$, both $E_C$ and $E_L$ monotonically increase, albeit with opposite curvatures. With increasing $c_{Mv}$, $S_C$ increases dramatically, while $S_L$ decreases slightly. Thus, with increasing $c_{Mv}$, the difference in $S_C$ and $S_L$ continuously decreases and the melting temperature increases; see Fig .5(a). However, latent heat per monomer $L_m$ goes through a maximum at $c_{Mv} \cong 0.03$ and then decreases with increasing $c_{Mv}$, when the drop in the entropy of melting exceeds the rise in the melting temperature.

**2. Solvent (or Free Volume) Density:** We now study the effect of solvent (or free volume) on properties at the melting point. To provide a reference for comparison, we fix $M_C$=2000 at the melting point; the rest of the parameters are as shown in Fig. 6. Flory[1] has obtained the following relation for the depression of the melting temperature due to the presence of low degree of polymerization diluents that is applicable to both polydisperse (these are *not* same as equilibrium polymerization considered here) and fixed length polymers:

$$1/T_{\rm M} - 1/T_{\rm M}^0 = \varepsilon[\phi_{\rm vL} + (\phi_{\rm mL} + \lambda)/M_{\rm L} - \delta\phi_{\rm vL}^2]/h_{\rm u}, \qquad (4.1)$$

where the notation has been modified slightly (recall that our temperature is scaled by $\varepsilon$, which explains its presence above). Here $M_L$ is the average DP in EL and $\phi_{\rm mL} = 1 - \phi_{\rm vL}$. For large $M_L$ and small $\phi_{\rm vL}$, $\lambda$ may be taken as a constant. The enthalpy of fusion per structural unit is $h_{\rm u}$, $\delta$ is related to the cohesive energy density, and $T_{\rm M}^0$ is the melting temperature of an infinite length polymer in the absence of diluents. The qualitative validity of Eq. (4.1) has been verified in simulations[18] and experiments.[54] In Fig. 6(a), we plot the inverse melting temperature against the solvent density in CR and EL. Although $M_L$ is changing [see Fig. 6(c)], we find that the linear relation given by Eq. (4.1) is roughly satisfied for EL (there is a slight positive curvature in $\phi_{\rm vL}$). For CR, however, although $M_C$ is fixed at 2000, $\phi_{\rm vC}$ is clearly *not* linear with $1/T_M$ if we consider the entire range of $\phi_{\rm vC}$. In Fig. 6(a), we also show the reduced pressure at the melting point, as a function of $\phi_{\rm vL}$. The reduced pressure behaves as we expect: it increases as $\phi_{\rm vL}$, and the corresponding $\phi_{\rm vC}$, decreases.

In Fig. 6(b), we show how $S_C$, $S_L$, $E_C$ and $E_L$ change with the melting temperature. We find that $S_C$, $S_L$, and $E_C$ are almost constant, but $E_L$ increases somewhat linearly with $T_M$. In Fig. 6(c), we show $M_L$ and $L_m$ at the melting point ($M_C$ fixed at 2000). We find that with increasing $T_M$, $L_m$ increases and reaches its asymptotic value at higher $T_M$, where the pressure is also higher; however, $M_L$ goes through a minimum.

**3. DP:** The first expression for the effect of DP on the equilibrium melting temperature was derived by Flory[1] by treating the end-groups as a different species with independent contributions to the free energy. In the absence of any diluent, a linear relation between the inverse melting temperature and the inverse average degree of polymerization was found, which was subsequently verified through experimental data[54] and simulations.[18]

We fix the pressure $P_r = 1$ at the melting point and plot $1/T_M$ against $1/M_C$ and $1/M_L$ in Fig. 7(a). We do not distinguish between the end-groups and the middle-groups ($c_{ME}$=0, $c_{Mv} = c_{Ev}$=0.01) in this case; the rest of the parameters are as shown. It is clear that a linear relation between $1/T_M$ and $1/M_L$ exists, in accordance with experimental data on fixed length polymers.[54] However, there is no such relationship for the crystal if we consider the entire range of $1/M_C$. As shown in Fig. 7(a), $T_M$ achieves its asymptotic value for high degree of polymerization as expected.

In Fig. 7(b), we show $S_C$, $S_L$, $E_C$ and $E_L$ at the melting point against $M_L$. We find that both $S_C$ and $S_L$ decrease with increasing $M_L$ and quickly achieve asymptotic values. The difference between $S_C$ and $S_L$ also stabilizes, and this causes $L_m$ to stabilize at high DP, as we see in Fig. 7(c). It has been suggested[3] that the latent heat per chain $L_P$ increases linearly with the chain length for fixed length polymers, especially at high degree of polymerizations. Thus, the latent heat per monomer $L_m$ reaches a constant value at high DP. Experimental data[3,54] support this view. This view is also consistent with our observation of stabilization of the latent heat per monomer $L_m$ at high DP. What we find from Fig. 7 is that the limiting value is achieved rather rapidly at $M_L \cong 600$, which is not too large. This is surprising.

With increasing $M_L$, $E_L$ increases while $E_C$ decreases; but both achieve asymptotic values at high DP. In Fig. 7(c), we show $\phi_{vC}$, $\phi_{vL}$ and $L_m$ against $M_L$. We find that $\phi_{vL}$ increases slightly but is almost constant. On the other hand $\phi_{vC}$ decreases with increasing $M_L$ and appears to go to zero or remain very small for high DP.

From Figs. 6(a) and 7(a), we notice that the relationships between $T_M$, and parameters such as the degree of polymerization and solvent density given in Flory's theory are satisfied qualitatively in equilibrium polymerization, provided we use the average DP and the solvent density of EL at the melting point in his theory.

A unique feature of our theory is that the end-group effects can be incorporated. We have investigated the effect of having attractive middle-group and end-group interactions ($c_{ME}<0$). We will only discuss our results here without presenting them, as the end-group effect is weak. For $c_{ME}<0$, $\phi_{vC}$ increases and $\phi_{vL}$ decreases; thus, the discontinuity in the free volume also decreases. Similarly the discontinuities in the average DP, the entropy and the energy per monomer also decrease. The effect on $T_M$ is too small to be discernible, while the latent heat per monomer slightly decreases. All these effects vanish for higher degree of polymerizations.

**4. Chain rigidity:** In the classical theory of Flory[2], chain rigidity is determined by the bending penalty $\varepsilon$. A similar definition for the chain rigidity has also been used by other authors.[4,52,55] In our theory however, there are three energy parameters $\varepsilon$, $\varepsilon'$, and $\varepsilon''$ that can all be related to the concept of rigidity. If we reduce $\varepsilon'$, the number of pairs of neighboring parallel bonds $N_{\rm p}$ will increase, which can be considered as a sign of higher chain rigidity. If $\varepsilon$ is increased, trans bonds will be favored, and this can also be considered as increasing the chain rigidity. In order to maximize $N_{\rm p}$ and minimize $N_{\rm g}$, we would like to have smaller $N_{\rm h}$, which will require a larger $\varepsilon''$. Thus, chain rigidity in our theory can be manipulated in more than one way, and we will consider the effects of $\varepsilon$, $\varepsilon'$ and $\varepsilon''$ separately to gain a better understanding of the effect of chain rigidity.

**(i) Effect of $\boldsymbol{\varepsilon}$:** In Fig. 8, we consider the effect of $\varepsilon$ on the melting properties by considering values of $a$ between 0.64 and 0.8. Since the temperature $T$, the reduced pressure, and the ratios $a$, $b$, $c_{\rm Mv}$, $c_{\rm Ev}$, and $c_{\rm ME}$ are all normalized by $\varepsilon$, care must be exercised in studying the effect of $\varepsilon$. Therefore, at the starting value, when $a$=0.64, we take $c_{\rm Mv}$=0.2, $c_{\rm ME}$=0.01, $c_{\rm Ev}$=0.3, $b$=0, $\mu_{\rm E}=-5$ and $P_{\rm r}=1$. Let $\varepsilon_0$ denote the value of $\varepsilon$ when $a$=0.64. When we consider any other value of $a$, the values of $c_{\rm Mv}$, $c_{\rm ME}$, $c_{\rm Ev}$, $b$, $\mu_{\rm E}$, and $P_{\rm r}$ are recalculated with the new $\varepsilon$ so that $w_{\rm Mv}$, $w_{\rm ME}$, $w_{\rm Ev}$, $w''$, $H$ at any given unscaled temperature $\widetilde{T}$ (recall that $\widetilde{T}$ is not scaled by $\varepsilon$) and the normal pressure $P$ are unchanged. This method ensures that only $\varepsilon$ is changing, while other quantities and the normal pressure are kept fixed.

In Fig. 8(a), we show how $T_{\rm M}$ and $L_{\rm m}$ change with $\varepsilon$. As there are different values of $\varepsilon$ for different choices of $a$, it will be difficult to compare the results if we report $T_{\rm M}$ which are normalized with different values of $\varepsilon$. Therefore, we have normalized the temperature $T$ by $\varepsilon_0$ everywhere when calculating $T_{\rm M}$ on the y-axis. We find that $T_{\rm M}$ decreases linearly with decreasing $\varepsilon$ (increasing $a$). The latent heat per monomer $L_m$ is also now normalized by $\varepsilon_0$ everywhere. It goes through a maximum and then decreases rapidly with increasing $\varepsilon$. This is an example of the interesting features observed in melting under equilibrium polymerization. For small $a$, the melting becomes continuous as evident from Fig. 8(b-c).

In the incompressible infinite chain length model,[19,20] it was shown that we need $a$>0 to yield a first order melting. For $a$=0, $b$=0, where that model reduces to the Flory model, the melting becomes continuous, in contradiction with the Flory calculation.[2] From Fig. 8, we find that to obtain first order melting, $a$ has to be large enough. In addition, we also need $a$<1; otherwise the ground state will not be the one with chains parallel to each other. For this model, when $b$=0, it appears that the range 0.6<$a$<1 would apply for real systems, where we expect a first order melting.

In Fig. 8(b), we show how $M_C$, $M_L$, $\phi_{vC}$ and $\phi_{vL}$ change with $\varepsilon$. We find that the change in $M_L$ with $\varepsilon$ is much smaller than the change in $M_C$. With increasing $\varepsilon$, $M_C$ continuously decreases until the melting becomes continuous for some $a$<0.64. At $a$=0.64 the difference between $M_C$ and $M_L$ is ≅ 200. The solvent density in CR increases with increasing $\varepsilon$ to ≅ 3% for $a$=0.64. These results on the changes in the solvent density with $\varepsilon$ are qualitatively consistent with those obtained by Flory for fixed length polymers.[2]

In Fig. 8(c), we show how $S_C$, $S_L$, $E_C$ and $E_L$ change with $\varepsilon$ at the melting point. Again, in this figure, the energies $E_C$ and $E_L$ are normalized by $\varepsilon_0$ everywhere for ease of comparison. We find that $S_C$, $S_L$, $E_C$ and $E_L$ all increase with increasing $\varepsilon$. However, $S_C$ and $E_C$ increase more rapidly than their counterparts in EL.

**(ii) Effect of $\varepsilon'$:** We have also investigated the effect of changing $\varepsilon'$ alone while keeping other parameters fixed. The results are shown in Fig. 9, where we have taken the range of $a$ between 0.64 and 0.88 but kept $c_{\rm Mv}$, $c_{\rm ME}$, $c_{\rm Ev}$, $b$, $\mu_{\rm E}$, and $P_{\rm r}=1$ fixed for all $a$. From Fig. 9(a), we find that $T_{\rm M}$ increases linearly, while $L_{\rm m}$ goes through a maximum and then decreases with decreasing $\varepsilon'$. From Fig. 9(b-c), we again find that the melting becomes continuous for small $a$.

The parameter $a$ plays an important role in the observance of first order melting. The energy between a pair of neighboring parallel bonds, $\varepsilon'$, has also been incorporated by other workers[15,18] in their model; however no information is available on the importance of this energy.

**(iii) Effect of $\varepsilon''$:** Gujrati et al[5-8] have shown that the hairpin turn excitations are important at low temperatures as they disorder the ground state. If these defects are suppressed, we expect the melting temperature to increase. In Fig. 10(a), we show how $T_{\rm M}$ and $L_{\rm m}$ vary with $b$ for $P_{\rm r}=1$. We find that both $T_{\rm M}$ and $L_{\rm m}$ increase with $b$ and the dependence is almost linear. In Fig. 10(b), we show the density of gauche bonds ($\phi_{\rm g}$) in CR and EL with $b$. We find that with increasing $b$, $\phi_{\rm g}$ in both CR and EL decrease; however, the difference is almost constant. The jump in $\phi_{\rm g}$ to higher values in going from CR to EL has also been studied by simulations.[13-15]

We note that decreasing $\varepsilon'$, increasing $\varepsilon$, and increasing $\varepsilon''$ all of which are related to the notion of increasing chain rigidity in our model have conflicting effects on the melting properties, such as $S_{\rm C}$, $S_{\rm L}$, $E_{\rm C}$, $E_{\rm L}$ and $L_{\rm m}$. Thus the different measures of chain rigidity may have competing effects on the melting properties and to properly characterize the effect of chain rigidity we will need to know the values of $\varepsilon$, $\varepsilon'$, and $\varepsilon''$.

In this section, we have studied the effect of various model parameters on the melting properties of equilibrium polymerization in a solution. To the best of our knowledge, no experimental data on the melting properties of equilibrium polymerization are available at present to make a more detailed comparison. Our results and the model should prove useful when such data do become available.

## V. Conclusions

We have considered a general model for crystallization under equilibrium polymerization in a solution. The model is solved on a Husimi cactus using recursion techniques. The CR phase is identified with a 2-cycle FP solution of the recursion relations, while the disordered equilibrium phase EL is identified with a 1-cycle solution. The Husimi cactus solution of the model is taken to be an approximate theory on the square lattice. The theory is thermodynamically consistent in the entire parameter space and provides a first principle basis to study polymers formed under conditions of equilibrium polymerization. The melting properties in equilibrium polymerization are studied with model parameters and qualitative agreement with some results from simulations and experimental data on fixed length polymers was observed. In particular, the effects of chain rigidity, monomer interactions, solvent quality and quantity, degree of polymerization, energy penalty for bends, parallel bonds, and hairpin turns (which are important at low temperatures) on properties such as the melting temperature, latent heat, and energy and entropy of fusion have been investigated. We find that the different measures of chain rigidity in our model may have competing effects on the melting properties, and therefore, care must be exercised in characterizing the effect of chain rigidity. To the best of our knowledge, such an explicit study of the relation between the melting properties in equilibrium polymerization and molecular parameters has not been conducted elsewhere. Our theory goes beyond the mean field approach of the classical theory and hence we believe our results will further the research in this area.

We would like to thank Andrea Corsi for his various discussions.

## Appendix

We shall use the following short-hand notation for simplicity

$X(\alpha) \equiv Z_m(\alpha)$, $Y(\alpha) \equiv Z_{m+1}(\alpha)$, $Z(\alpha) \equiv Z_{m+2}(\alpha)$, $u \equiv w_{\mathrm{ME}}$, $z \equiv w_{\mathrm{Ev}}$, and $y \equiv w_{\mathrm{Mv}}$.

### A. Crystalline phase: 2-Cycle Scheme

The PPFs $Z_m(L)$ and $Z_m(R)$ can be broken into contributions from trans and gauche states. Therefore we have,

$$X(L) = X(L_{\mathrm{t}}) + X(L_{\mathrm{g}}),\ X(R) = X(R_{\mathrm{t}}) + X(R_{\mathrm{g}}).$$

Owing to the symmetry of $L$ and $R$ states, we also have the following identities:

$$X(R_{\mathrm{t}}) = X(L_{\mathrm{g}})/w,\ X(R_{\mathrm{g}}) = wX(L_{\mathrm{t}}).$$

We obtain the following seven independent recursion relations:

$$\begin{aligned}
X(L_{\mathrm{t}}) = \eta_{\mathrm{M}}[&uZ(I)Y(I)Y(E_b) + uy^2Y(E_b)Y(v)Z(I) + zyY(E_b)Z(v)[Y(I)+Y(v)] \\
&+ u^3Y(E_b)Z(I)Y(E_a) + z^2uY(E_b)Z(v)Y(E_a) + uY(E_b)Y(I)Z(E_a) \\
&+ zyY(E_b)Y(v)Z(E_a) + uZ(E_a)Y(E_a)Y(E_b) + Z(I)Y(I)Y(R) + y^2Y(v)Z(I)Y(R) \\
&+ y^2Z(v)Y(I)Y(R) + y^2Z(v)Y(v)Y(R) + u^2Y(E_a)Z(I)Y(R) + yzuZ(v)Y(E_a)Y(R) \\
&+ u^2Z(E_a)Y(I)Y(R) + uzyZ(E_a)Y(v)Y(R) + u^2Z(E_a)Y(E_a)Y(R) \\
&+ u^2w'Y^2(E_b)Z(L) + w'Z(E_b)Y(E_b)Y(R) + 2uw'Y(E_b)Z(L)Y(R) \\
&+ uw'Z(E_b)Y^2(E_b) + uw'Z(E_b)Y^2(R) + w'Z(L)Y^2(R) + u^2w'Z(E_b)Y(E_b)Y(R) \\
&+ wZ(R)Y(O)Y(I) + wy^2Y(O)Y(v)Z(R) + wu^2Y(E_a)Y(O)Z(R) \\
&+ wuZ(E_b)Y(O)Y(I) + wzyZ(E_b)Y(v)Y(O) + wuY(E_a)Z(E_b)Y(O) \\
&+ w''w^2w'Z(O)Y(O)[uY(E_b) + Y(R)]],
\end{aligned}$$

$$
\begin{aligned}
X(L_{\mathrm{g}}) = \eta_{\mathrm{M}}[w[uZ(I)Y(I)Y(E_b) + uy^2Y(E_b)Y(v)Z(I) + zyY(E_b)Z(v)[Y(I) + Y(v)] \\
+ u^3Y(E_b)Y(E_a)Z(I) + z^2uY(E_b)Y(E_a)Z(v) + uY(E_b)Y(I)Z(E_a) \\
+ zyY(E_b)Y(v)Z(E_a) + uY(E_b)Y(E_a)Z(E_a)] + wY(L)[Y(I)Z(I) + y^2Y(v)Z(I) \\
+ y^2Y(I)Z(v) + y^2Z(v)Y(v) + u^2Z(I)Y(E_a) + yzuZ(v)Y(E_a) + u^2Z(E_a)Y(I) \\
+ uzyZ(E_a)Y(v) + u^2Z(E_a)Y(E_a)] + w[u^2w'Y^2(E_b)Z(R) + w'Z(E_b)Y(E_b)Y(L) \\
+ uw'Z(R)Y(L)Y(E_b) + uw'Z(E_b)Y^2(E_b) + uw'Z(R)Y(E_b)Y(L) \\
+ uw'Z(E_b)Y^2(L) + w'Z(R)Y^2(L) + u^2w'Z(E_b)Y(E_b)Y(L)] \\
+ w[wZ(L)Y(O)Y(I) + wy^2Z(L)Y(v)Y(O) + wu^2Z(L)Y(E_a)Y(O) \\
+ wuZ(E_b)Y(O)Y(I) + wzyZ(E_b)Y(O)Y(v) + wuZ(E_b)Y(E_a)Y(O) \\
+ w''w^2w'Z(O)Y(O)[uY(E_b) + Y(L)]]],
\end{aligned}
$$

$$
\begin{aligned}
X(v) = Z(v)Y^2(v) + 2z^2Z(v)Y(v)Y(E_a) + z^2Y^2(v)Z(E_a) + 2y^2Z(v)Y(v)Y(I) \\
+ y^2Y^2(v)Z(I) + 2z^2Y(v)Y(E_a)Z(E_a) + 2uyzZ(E_a)Y(v)Y(I) + 2uzyY(E_a)Y(v)Z(I) \\
+ 2y^2Y(v)Y(I)Z(I) + z^4Z(v)Y^2(E_a) + 2z^2y^2Y(E_a)Z(v)Y(I) + y^4Z(v)Y^2(I) \\
+ yzY(v)Y(E_b)[Z(R) + Z(L)] + yzY(v)Z(E_b)[Y(L) + Y(R)] \\
+ y^2Y(v)[Z(R)Y(L) + Y(R)Z(L)] + 2z^2Y(v)Y(E_b)Z(E_b) + z^2Y^2(E_a)Z(E_a) \\
+ 2uzyZ(E_a)Y(E_a)Y(I) + u^2z^2Y^2(E_a)Z(I) + 2uzyZ(I)Y(I)Y(E_a) \\
+ u^2y^2Z(E_a)Y^2(I) + y^2Z(I)Y^2(I) + z^2uY(E_b)Y(E_a)[Z(R) + Z(L)] \\
+ zyY(E_b)Y(I)[Z(R) + Z(L)] + yzZ(E_b)Y(E_a)[Y(L) + Y(R)] \\
+ uy^2Z(E_b)Y(I)[Y(L) + Y(R)] + uzyY(E_a)[Z(R)Y(L) + Y(R)Z(L)] \\
+ y^2Y(I)[Z(R)Y(L) + Y(R)Z(L)] + 2z^2Z(E_b)Y(E_b)Y(E_a) + 2uzyZ(E_b)Y(E_b)Y(I) \\
+ y^2wZ(O)Y(L)Y(R) + z^2wY^2(E_b)Z(O) + zywZ(O)Y(E_b)[Y(L) + Y(R)],
\end{aligned}
$$

$$
\begin{aligned}
X(E_b) = H[&Z(E_a)Y^2(E_a) + 2Z(E_b)Y(E_b)Y(E_a) + 2z^2Y(v)Y(E_a)Z(E_a) + z^2Z(v)Y^2(E_a) \\
&+ 2u^2Y(E_\mathrm{a})Y(I)Z(E_a) + u^2Y^2(E_a)Z(I) + 2z^2Y(v)Z(E_b)Y(E_b) + 2u^2Z(E_b)Y(E_b)Y(I) \\
&+ uY(E_b)Y(E_a)[Z(R) + Z(L)] + uZ(E_b)Y(E_a)[Y(L) + Y(R)] + wY^2(E_b)Z(O) \\
&+ 2z^2Y(E_a)Y(v)Z(v) + 2zyuY(E_a)[Z(v)Y(I) + Y(v)Z(I)] + 2u^2Y(E_a)Y(I)Z(I) \\
&+ z^4Z(E_a)Y^2(v) + 2z^2u^2Y(v)Z(E_a)Y(I) + u^4Z(E_a)Y^2(I) + uY(I)Y(E_b)[Z(R) + Z(L)] \\
&+ zyY(E_b)Y(v)[Z(R) + Z(L)] + u^3Y(I)Z(E_b)[Y(L) + Y(R)] \\
&+ z^2uZ(E_b)Y(v)[Y(L) + Y(R)] + u^2Y(E_a)[Y(L)Z(R) + Z(L)Y(R)] \\
&+ wuY(E_b)Z(O)[Y(R) + Y(L)] + u^2Z(I)Y^2(I) + 2uzyY(v)Y(I)Z(I) + z^2u^2Z(v)Y^2(I) \\
&+ 2zuyZ(v)Y(v)Y(I) + z^2y^2Y^2(v)Z(I) + z^2Z(v)Y^2(v) + u^2Y(I)[Y(L)Z(R) + Z(L)Y(R)] \\
&+ yzuY(v)[Y(L)Z(R) + Z(L)Y(R)] + u^2wY(L)Y(R)Z(O)],
\end{aligned}
$$

$$
\begin{aligned}
X(O) = \eta_\mathrm{M}[&Z(I)Y^2(I) + 2y^2Z(I)Y(I)Y(v) + y^2Y^2(I)Z(v) + 2u^2Z(I)Y(I)Y(E_a) \\
&+ u^2Y^2(I)Z(E_a) + 2yzuY(I)Y(v)Z(E_a) + 2yzuY(I)Z(v)Y(E_a) \\
&+ 2y^2u^2Z(I)Y(E_a)Y(v) + 2y^2Z(v)Y(v)Y(I) + 2u^2Y(I)Y(E_a)Z(E_a) + y^4Y^2(v)Z(I) \\
&+ u^4Z(I)Y^2(E_a) + Y(I)[Z(L)Y(R) + Y(L)Z(R)] + uY(I)[Y(E_b)Z(L) + Y(R)Z(E_b) \\
&+ Y(L)Z(E_b) + Y(E_b)Z(R)] + 2u^2Y(I)Y(E_b)Z(E_b) + y^2Y^2(v)Z(v) \\
&+ 2yzuY(E_a)Y(v)Z(v) + y^2z^2Y^2(v)Z(E_a) + 2yzuY(v)Y(E_a)Z(E_a) \\
&+ z^2u^2Y^2(E_a)Z(v) + u^2Y^2(E_a)Z(E_a) + u^3Z(L)Y(E_b)Y(E_a) + y^2uY(v)Y(E_b)Z(L) \\
&+ u^3Z(R)Y(E_a)Y(E_b) + y^2uY(v)Y(E_b)Z(R) + uY(R)Z(E_b)Y(E_a) \\
&+ yzY(v)Z(E_b)[Y(R) + Y(L)] + uY(L)Z(E_b)Y(E_a) \\
&+ u^2Y(E_a)[Z(L)Y(R) + Y(L)Z(R)] + y^2Y(v)[Z(L)Y(R) + Y(L)Z(R)] \\
&+ 2u^2Z(E_b)Y(E_b)Y(E_a) + 2yzuZ(E_b)Y(E_b)Y(v) + wY(L)Y(R)Z(O) \\
&+ wuY(E_b)Z(O)[Y(L) + Y(R)] + wu^2Y^2(E_b)Z(O)],
\end{aligned}
$$

$$
\begin{aligned}
X(I) = \eta_\mathrm{M}[&wY(L)Y(R)Z(I) + wy^2Y(L)Y(R)Z(v) + wu^2Y(L)Y(R)Z(E_a) \\
&+ wuY(E_b)Z(I)[Y(R) + Y(L)] + wzyY(E_b)Z(v)[Y(R) + Y(L)] \\
&+ wuY(E_b)Z(E_a)[Y(R) + Y(L)] + wu^2Z(I)Y^2(E_b) + wz^2Z(v)Y^2(E_b) + wZ(E_a)Y^2(E_b) \\
&+ w''w^2w'[2Y(O)Y(E_b)Z(E_b) + uY(O)Y(E_b)[Z(R) + Z(L)] + uY(O)Z(E_\mathrm{b})[Y(R) + Y(L)] \\
&+ Y(O)[Y(R)Z(L) + Z(R)Y(L)]]],
\end{aligned}
$$

$$
\begin{aligned}
X(E_a) = H[&2Z(E_a)Y(E_a)Y(E_b) + 2w'Z(E_b)Y^2(E_b) + 2z^2Y(E_b)Z(E_a)Y(v) \\
&+ 2u^2Z(E_a)Y(E_b)Y(I) + 2z^2Y(E_a)Y(E_b)Z(v) + 2u^2Y(E_b)Y(E_a)Z(I) \\
&+ uZ(E_a)Y(E_a)[Y(L) + Y(R)] + w'uY^2(E_b)[Z(R) + Z(L)] \\
&+ 2w'uZ(E_b)Y(E_b)[Y(L) + Y(R)] + 2wZ(E_b)Y(E_a)Y(O) + 2z^2Y(E_b)Y(v)Z(v) \\
&+ 2zyuY(E_b)[Z(I)Y(v) + Y(I)Z(v)] + 2u^2Y(E_b)Y(I)Z(I) \\
&+ uY(E_a)Z(I)[Y(R) + Y(L)] + zyY(E_a)Z(v)[Y(R) + Y(L)] \\
&+ u^3Z(E_a)Y(I)[Y(R) + Y(L)] + z^2uZ(E_a)Y(v)[Y(R) + Y(L)] \\
&+ u^2w'Y(E_b)[Y(R)Z(L) + Z(R)Y(L)] + w'Y(E_b)[Z(L)Y(R) + Y(L)Z(R)] \\
&+ w'u^2Z(E_b)[Y^2(R) + Y^2(L)] + uwY(E_a)Y(O)[Z(L) + Z(R)] \\
&+ 2w^2w'w''Y(E_b)Y(O)Z(O) + uZ(I)Y(I)[Y(R) + Y(L)] \\
&+ zyY(v)Z(I)[Y(R) + Y(L)] + uy^2Z(v)Y(I)[Y(R) + Y(L)] \\
&+ zyZ(v)Y(v)[Y(R) + Y(L)] + uw'[Z(L)Y^2(R) + Z(R)Y^2(L)] \\
&+ wuY(O)Y(I)[Z(R) + Z(L)] + wzyY(v)Y(O)[Z(R) + Z(L)] \\
&+ w^2uw'w''Z(O)Y(O)[Y(R) + Y(L)]].
\end{aligned}
$$

The ratios *l* and *r* can be broken into contributions from trans and gauche states at the fixed point (in CR and EL):

$$
l = l_{\mathrm{t}} + l_{\mathrm{g}}, \; r = r_{\mathrm{t}} + r_{\mathrm{g}}.
$$

In CR, they have the following relationships: $r_{\mathrm{t}} = l_{\mathrm{g}} / w$, $r_{\mathrm{g}} = wl_{\mathrm{t}}$. We introduce the quantities $Q_{\mathrm{LR}}$ and $Q_2$, which are given by the relations,

$$
\Psi_m = \Psi_{m+1}^2 \Psi_{m+2} Q_{\mathrm{LR}}, \; Z = \Psi_0^2 Q_2, \; \Psi_m \equiv Z_m(R) + Z_m(L).
$$

As there are two types of sites A and B, we obtain two expressions for $Q_{\mathrm{LR}}$ and $Q_2$. We present the equations for $Q_{\mathrm{LR}}$ and $Q_2$ below. For ease of writing, we denote the quantities evaluated on site A by a bar and those on site B with a hat. We get,

$$
\overline{Q}_2 = (2\overline{o}\,\overline{i} + \overline{l}_{\mathrm{t}}^2 + \overline{r}_{\mathrm{t}}^2 + 2\overline{r}_{\mathrm{g}}\overline{l}_{\mathrm{g}} / w) / \eta_{\mathrm{M}} + \overline{s}^2 + 2\overline{e}_a\overline{e}_b / H,
$$

$$
\begin{aligned}
\overline{Q}_{\mathrm{LR}} = \eta_{\mathrm{M}}(1+w)[& u\bar{i}\hat{i}\hat{e}_b + uy^2\hat{e}_b\hat{s}\bar{i} + zy\hat{e}_b\bar{s}\hat{i} + zy\bar{s}\hat{s}\hat{e}_b + u^3\hat{e}_b\bar{i}\hat{e}_a + z^2u\hat{e}_b\bar{s}\hat{e}_a + u\hat{e}_b\hat{i}\bar{e}_a \\
& + zy\hat{e}_b\hat{s}\bar{e}_a + u\bar{e}_a\hat{e}_a\hat{e}_b + \bar{i}\hat{i}\hat{r} + y^2\hat{s}\bar{i}\hat{r} + y^2\bar{s}\hat{i}\hat{r} + y^2\bar{s}\hat{s}\hat{r} + u^2\hat{e}_a\bar{i}\hat{r} + yzu\bar{s}\hat{e}_a\hat{r} \\
& + u^2\bar{e}_a\hat{i}\hat{r} + uzy\bar{e}_a\hat{s}\hat{r} + u^2\bar{e}_a\hat{e}_a\hat{r} + u^2w'\hat{e}_b^2\bar{l} + w'\bar{e}_b\hat{e}_b\hat{r} + uw'\hat{e}_b\bar{l}\hat{r} + uw'\bar{e}_b\hat{e}_b^2 \\
& + uw'\hat{e}_b\bar{l}\hat{r} + uw'\bar{e}_b\hat{r}^2 + w'\bar{l}\hat{r}^2 + u^2w'\bar{e}_b\hat{e}_b\hat{r} + w\bar{r}\hat{o}\hat{i} + wy^2\hat{o}\hat{s}\bar{r} + wu^2\hat{e}_a\hat{o}\bar{r} \\
& + wu\bar{e}_b\hat{o}\hat{i} + wzy\bar{e}_b\hat{s}\hat{o} + wu\hat{e}_a\bar{e}_b\hat{o} + w^2w'u\bar{o}\hat{o}\hat{e}_b + w^2w'\bar{o}\hat{o}\hat{r} + u\bar{i}\hat{i}\hat{e}_b \\
& + uy^2\hat{e}_b\hat{s}\bar{i} + zy\hat{e}_b\bar{s}\hat{i} + zy\bar{s}\hat{s}\hat{e}_b + u^3\hat{e}_b\hat{e}_a\bar{i} + z^2u\hat{e}_b\hat{e}_a\bar{s} + u\hat{e}_b\hat{i}\bar{e}_a + zy\hat{e}_b\hat{s}\bar{e}_a \\
& + u\hat{e}_b\hat{e}_a\bar{e}_a + \hat{l}(\bar{i}\hat{i} + y^2\hat{s}\bar{i} + y^2\bar{s}\hat{i} + y^2\bar{s}\hat{s} + u^2\hat{e}_a\bar{i} + yzu\bar{s}\hat{e}_a + u^2\bar{e}_a\hat{i} + uzy\bar{e}_a\hat{s} \\
& + u^2\bar{e}_a\hat{e}_a) + u^2w'\hat{e}_b^2\bar{r} + w'\bar{e}_b\hat{e}_b\hat{l} + uw'\hat{e}_b\bar{r}\hat{l} + uw'\bar{e}_b\hat{e}_b^2 + uw'\hat{e}_b\hat{l}\bar{r} + uw'\bar{e}_b\hat{l}^2 \\
& + w'\bar{r}\hat{l}^2 + u^2w'\bar{e}_b\hat{e}_b\hat{l} + w\bar{l}\hat{o}\hat{i} + wy^2\hat{o}\hat{s}\bar{l} + wu^2\hat{e}_a\hat{o}\bar{l} + wu\bar{e}_b\hat{o}\hat{i} + wzy\bar{e}_b\hat{s}\hat{o} \\
& + wu\hat{e}_a\bar{e}_b\hat{o} + w^2w'u\bar{o}\hat{o}\hat{e}_b + w^2w'\bar{o}\hat{o}\hat{l}].
\end{aligned}
$$

To obtain $\hat{Q}_2$ and $\hat{Q}_{\mathrm{LR}}$, we simply replace the bar and hat on *i*, *o*, *s*, *r*, *l*, $e_a$, and $e_b$ in the above equations with a hat and bar respectively.

The adimensional pressure,

$$z_0 \equiv \beta P v_0 = 0.5\ln(\overline{Q}_{\mathrm{LR}} / \hat{Q}_2) = 0.5\ln(\hat{Q}_{\mathrm{LR}} / \overline{Q}_2).$$

Next we present the equations for various densities, which can be applied for both sites A and B. Use of either kind of site gives identical results. The density of the solvent, monomer, internal and external bends, end-groups, left-trans, right-trans, left-gauche, right-gauche, left and right turns, and gauche bonds are given by,

$\phi_{\mathrm{v}} = \bar{s}^2 / \overline{Q}_2$, $\phi_{\mathrm{m}} = 1 - \phi_{\mathrm{v}}$, $\phi_{\mathrm{i}} = \phi_{\mathrm{o}} = 2\bar{o}\bar{i} / (\eta_{\mathrm{M}}\overline{Q}_2)$, $\phi_{\mathrm{e}} = 2\bar{e}_a\bar{e}_b / (H\overline{Q}_2)$,

$\phi_{l_{\mathrm{t}}} = \bar{l}_{\mathrm{t}}^2 / (\eta_{\mathrm{M}}\overline{Q}_2)$, $\phi_{r_{\mathrm{t}}} = \bar{r}_{\mathrm{t}}^2 / (\eta_{\mathrm{M}}\overline{Q}_2)$, $\phi_{l_{\mathrm{g}}} = \phi_{r_{\mathrm{g}}} = \bar{r}_{\mathrm{g}}\bar{l}_{\mathrm{g}} / (w\eta_{\mathrm{M}}\overline{Q}_2)$, $\phi_l = \phi_{l_{\mathrm{t}}} + \phi_{l_{\mathrm{g}}}$,

$\phi_r = \phi_{r_{\mathrm{t}}} + \phi_{r_{\mathrm{g}}}$, $\phi_{\mathrm{g}} = \phi_{\mathrm{i}} + \phi_{r_{\mathrm{g}}} + \phi_{l_{\mathrm{g}}}$.

The corresponding equations for site B (with a hat) can be obtained by replacing the bar with a hat above. The average degree of polymerization $M = \phi_{\mathrm{m}} / \phi_{\mathrm{n}}$, and the number density $\phi_{\mathrm{n}} = \phi_{\mathrm{e}} / 2$.

## B. Amorphous phase: 1-Cycle Scheme

The recursion relations for the crystal also yield the solution for the amorphous phase when we set $Z(\alpha) \equiv Y(\alpha)$. In this case we also get other simple relations:

$$l_{\mathrm{t}} = 0.5/(1+w)\,,\ l_{\mathrm{g}} = wl_{\mathrm{t}}\,,\ r_{\mathrm{t}} = l_{\mathrm{t}}\,,\ r_{\mathrm{g}} = l_{\mathrm{g}}$$

## C. Statistical weight of each configuration in Fig. 4

(a) $\eta_{\mathrm{M}} wY(L)Y(R)Z(I)$ [for state *I* at (*m*+2) level], $\eta_{\mathrm{M}} wy^2 Y(L)Y(R)Z(v)$ [for state *v* at (*m*+2) level], and $\eta_{\mathrm{M}} wu^2 Y(L)Y(R)Z(E_{\mathrm{a}})$ [for state $E_{\mathrm{a}}$ at (*m*+2) level], (b) $\eta_{\mathrm{M}} wuY(E_{\mathrm{b}})Y(R)Z(I)$ [for state *I* at (*m*+2) level], $\eta_{\mathrm{M}} wzyY(E_{\mathrm{b}})Y(R)Z(v)$ [for state *v* at (*m*+2) level], and $\eta_{\mathrm{M}} wuY(E_{\mathrm{b}})Y(R)Z(E_{\mathrm{a}})$ [for state $E_{\mathrm{a}}$ at (*m*+2) level],

(c) $\eta_{\mathrm{M}} wuY(E_{\mathrm{b}})Y(L)Z(I)$ [for state *I* at (*m*+2) level], $\eta_{\mathrm{M}} wzyY(E_{\mathrm{b}})Y(L)Z(v)$ [for state *v* at (*m*+2) level], and $\eta_{\mathrm{M}} wuY(E_{\mathrm{b}})Y(L)Z(E_{\mathrm{a}})$ [for state $E_{\mathrm{a}}$ at (*m*+2) level],

(d) $\eta_{\mathrm{M}} wu^2 Y^2(E_{\mathrm{b}})Z(I)$ [for state *I* at (*m*+2) level], $\eta_{\mathrm{M}} wz^2 Y^2(E_{\mathrm{b}})Z(v)$ [for state *v* at (*m*+2) level], and $\eta_{\mathrm{M}} wY^2(E_{\mathrm{b}})Z(E_{\mathrm{a}})$ [for state $E_{\mathrm{a}}$ at (*m*+2) level],

(e) $\eta_{\mathrm{M}} w^2 w' w'' Y(O)Y(E_{\mathrm{b}})Z(E_{\mathrm{b}})$, (f) $\eta_{\mathrm{M}} w^2 w' w'' Y(O)Y(E_{\mathrm{b}})Z(E_{\mathrm{b}})$,

(g) $\eta_{\mathrm{M}} w^2 w' w'' uY(O)Y(E_{\mathrm{b}})Z(L)$, (h) $\eta_{\mathrm{M}} w^2 w' w'' uY(O)Y(E_{\mathrm{b}})Z(R)$,

(i) $\eta_{\mathrm{M}} w^2 w' w'' uY(O)Y(R)Z(E_{\mathrm{b}})$, (j) $\eta_{\mathrm{M}} w^2 w' w'' uY(O)Y(L)Z(E_{\mathrm{b}})$,

(k) $\eta_{\mathrm{M}} w^2 w' w'' Y(O)Y(R)Z(L)$, and (l) $\eta_{\mathrm{M}} w^2 w' w'' Y(O)Y(L)Z(R)$.

## Figure captions

Fig. 1. Husimi cactus and various levels. The level $m$ increases as we move away from the origin $m$=0. The numbers denote the indexing used for the crystalline phase. If the bottom vertex of a square is of level $m$, the two intermediate vertices are of level ($m$+1), and the top vertex is of level ($m$+2). Even and odd numbered sites are labeled as two separate kinds of sites, A or B.

Fig. 2. The five distinct states in a square cell (a) σ=0; no polymer bond (b) σ=1; two neighboring sites occupied by one bond (c) σ=g; two neighboring bonds making a bend (d) σ=p; two parallel bonds (e) σ=h; three bonds forming a hairpin turn.

Fig. 3. The seven states at a site and the possible ground states. The end-groups are denoted by a cross (×) and the solvent or void by an empty circle (○), the remaining sites are covered by the middle-groups. The filled dots (●) denote the corner of $\Sigma$, with $\Sigma'$ across it. We show the sequence of dots for configurations 1 and 4 of the walk.

Fig. 4. Distinct configurations of the square with the level $m$ in state $I$. The statistical weight of each configuration is given in section C of the Appendix.

Fig. 5. Effect of $c_{\mathrm{Mv}}$ at fixed reduced pressure $P_{\mathrm{r}}=1$, on (a) the melting temperature and the latent heat per monomer, (b) degree of polymerization and solvent density in the crystal and liquid at the melting point, (c) entropy and energy per monomer in the crystal and liquid. The other parameters are fixed as follows: $a$=0.7, $c_{\mathrm{ME}}$=0.01, $c_{\mathrm{Ev}}$=0.04, $\mu_{\mathrm{E}}$= −5, $b$=0.

Fig. 6. Effect of solvent density on various properties, when the degree of polymerization in the crystal at the melting point is fixed at 2000. (a) The inverse melting temperature with the solvent density in crystal and liquid. We also show the reduced pressure at the melting point as a function of the solvent density in the liquid. (b) Entropy and energy per monomer in the crystal and liquid with the melting temperature corresponding to different pressures at the melting point. (c) The degree of polymerization in the liquid and the latent heat per monomer with the melting temperature. The other parameters are fixed as follows: $a$=0.8, $c_{\mathrm{Mv}}$=0.01, $c_{\mathrm{Ev}}$=0.01, $c_{\mathrm{ME}}$=0.01, $b$=0.

Fig. 7. Effect of degree of polymerization on various properties for fixed pressure $P_{\mathrm{r}}=1$ at the melting point. (a) The inverse melting temperature, with the inverse degree of polymerization of the liquid and the crystal. (b) Entropy and energy per monomer in the crystal and the liquid with the degree of polymerization of the liquid at the melting point. (c) The solvent density in the crystal and liquid, and the latent heat per monomer with the degree of polymerization of the liquid. The other parameters are fixed as follows: $a$=0.8, $c_{\mathrm{Mv}}$=0.01, $c_{\mathrm{ME}}$=0, $c_{\mathrm{Ev}}$=0.01, $b$=0.

Fig. 8. Effect of the bending penalty $\varepsilon$ on the melting properties for fixed pressure. The melting temperature $T_{\mathrm{M}}$, the energies $E_{\mathrm{C}}$, $E_{\mathrm{L}}$, and the latent heat per monomer $L_{\mathrm{m}}$ are normalized by $\varepsilon_0$, which is the value of $\varepsilon$ when $a$=0.64. When $a$=0.64 the other parameters are $c_{\mathrm{Mv}}$=0.2, $c_{\mathrm{ME}}$=0.01, $c_{\mathrm{Ev}}$=0.3, $b$=0, $\mu_{\mathrm{E}}=-5$ and $P_{\mathrm{r}}=1$. When $a$≠0.64 these parameters are recalculated with the new $\varepsilon$ such that $w_{\mathrm{Mv}}$, $w_{\mathrm{ME}}$, $w_{\mathrm{Ev}}$, $w''$, $H$ (at any given $\widetilde{T}$) and the pressure $P$ are unchanged. We show (a) Melting temperature and latent heat per monomer. (b) Degree of polymerization and solvent density in the crystal

and liquid. (c) Entropy and energy per monomer in the crystal and liquid at the melting point.

Fig. 9. Effect of the parallel bond energy $\varepsilon'$ on the melting properties for fixed pressure $P_r = 1$. The other parameters are fixed as follows: $c_{Mv}$=0.2, $c_{ME}$=0.01, $c_{Ev}$=0.3, $b$=0, $\mu_E = -5$. We show (a) Melting temperature and latent heat per monomer. (b) Degree of polymerization and solvent density in the crystal and liquid. (c) Entropy and energy per monomer in the crystal and liquid at the melting point.

Fig. 10. Effect of energy penalty for hairpin turns $\varepsilon''$ on the melting properties for fixed pressure $P_r = 1$. The other parameters are fixed as follows: $a$=0.8, $c_{Mv}$=0.2, $c_{ME} = -0.05$, $c_{Ev}$=0.1, $\mu_E = -5$. We show (a) Melting temperature and latent heat per monomer. (b) Density of gauche bonds in the crystal and the liquid.

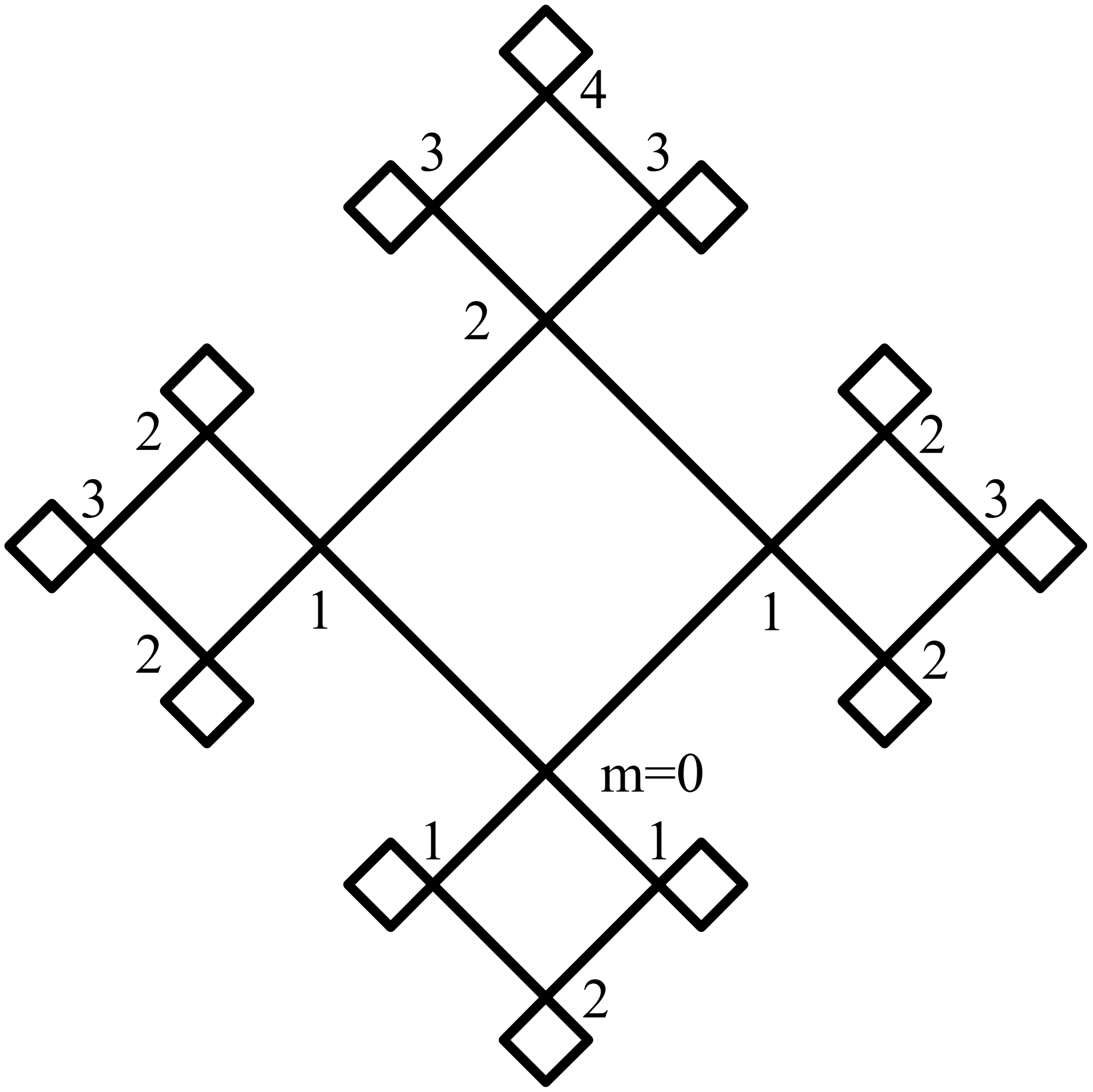

Figure 1

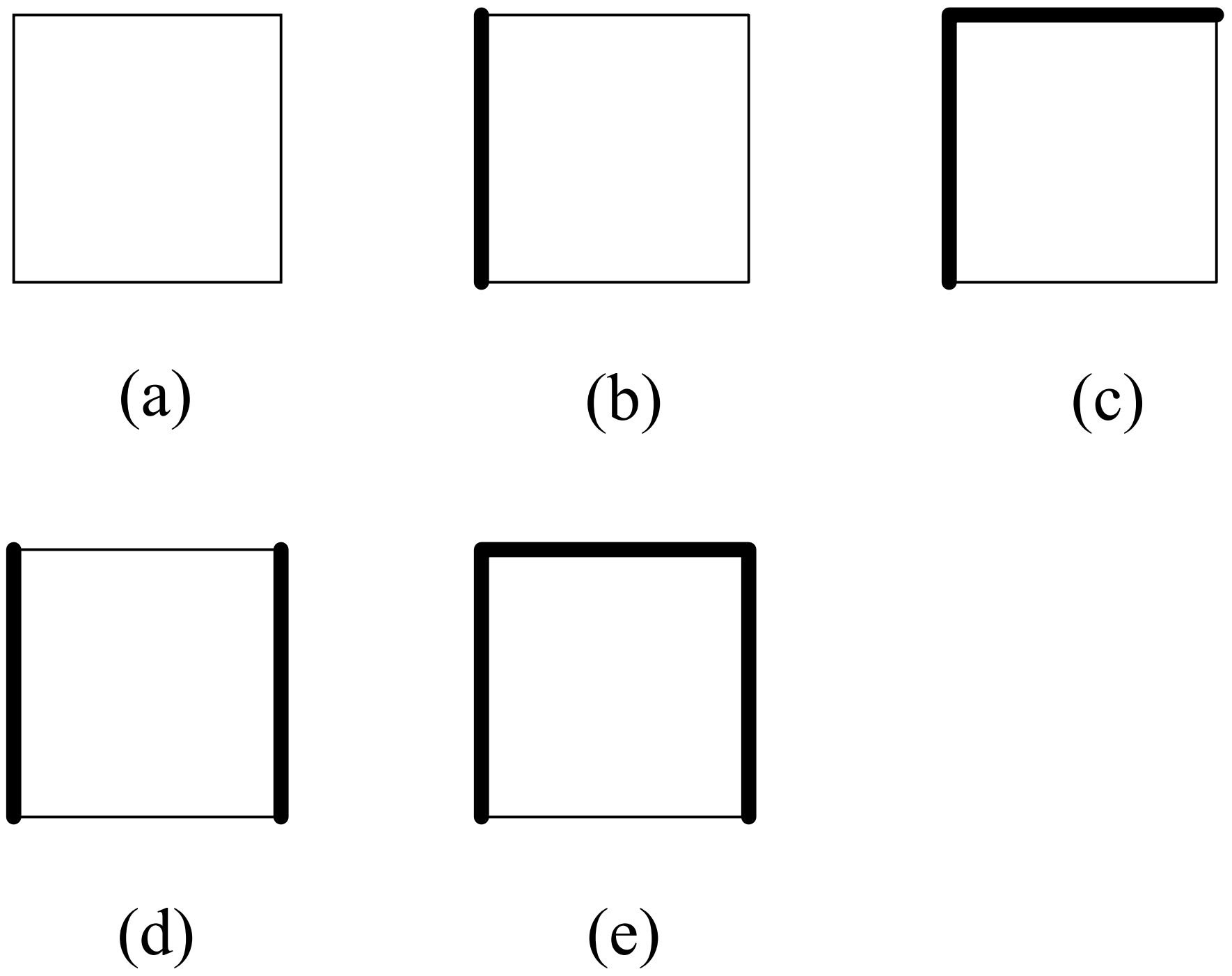

Figure 2

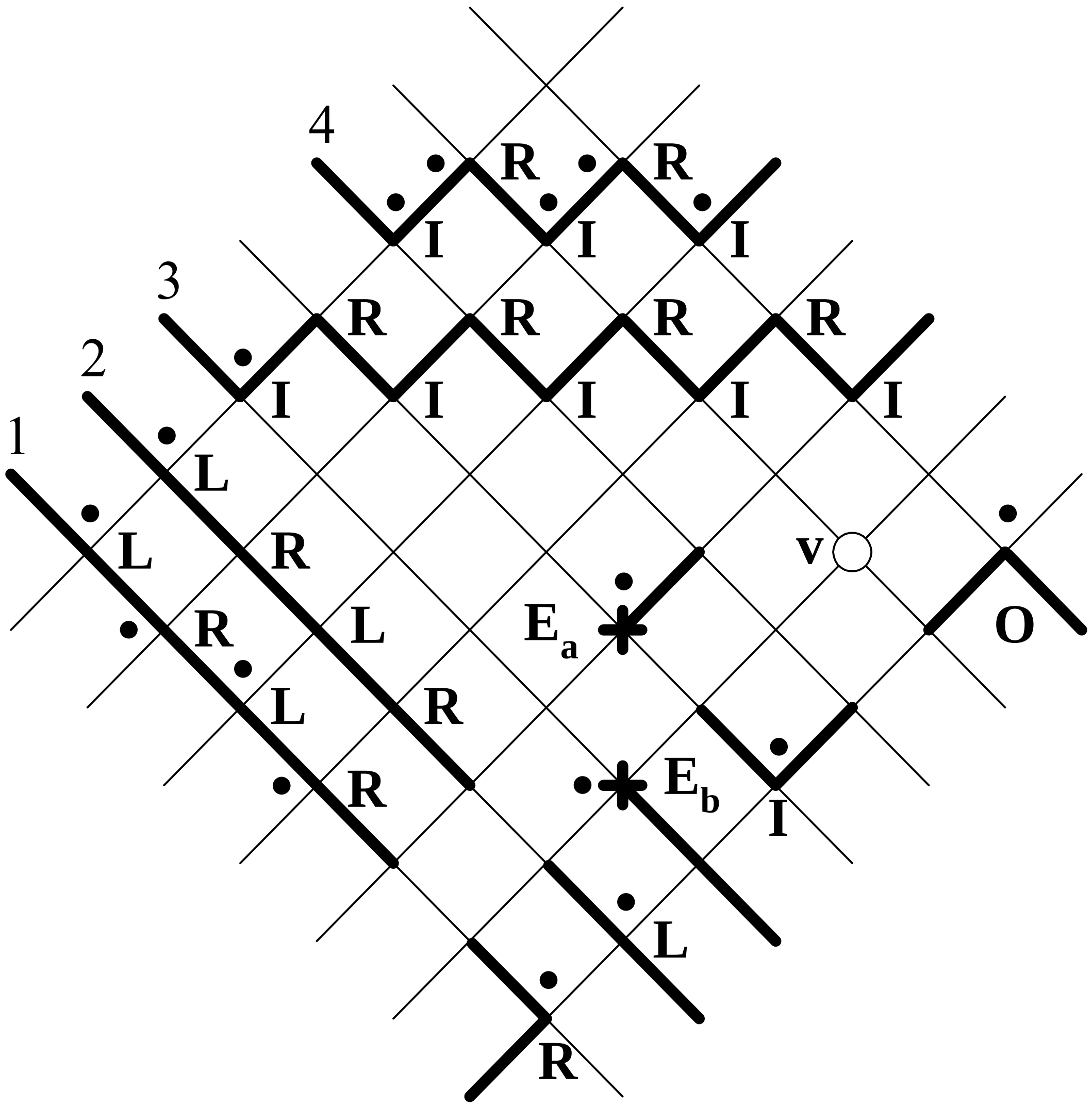

Figure 3

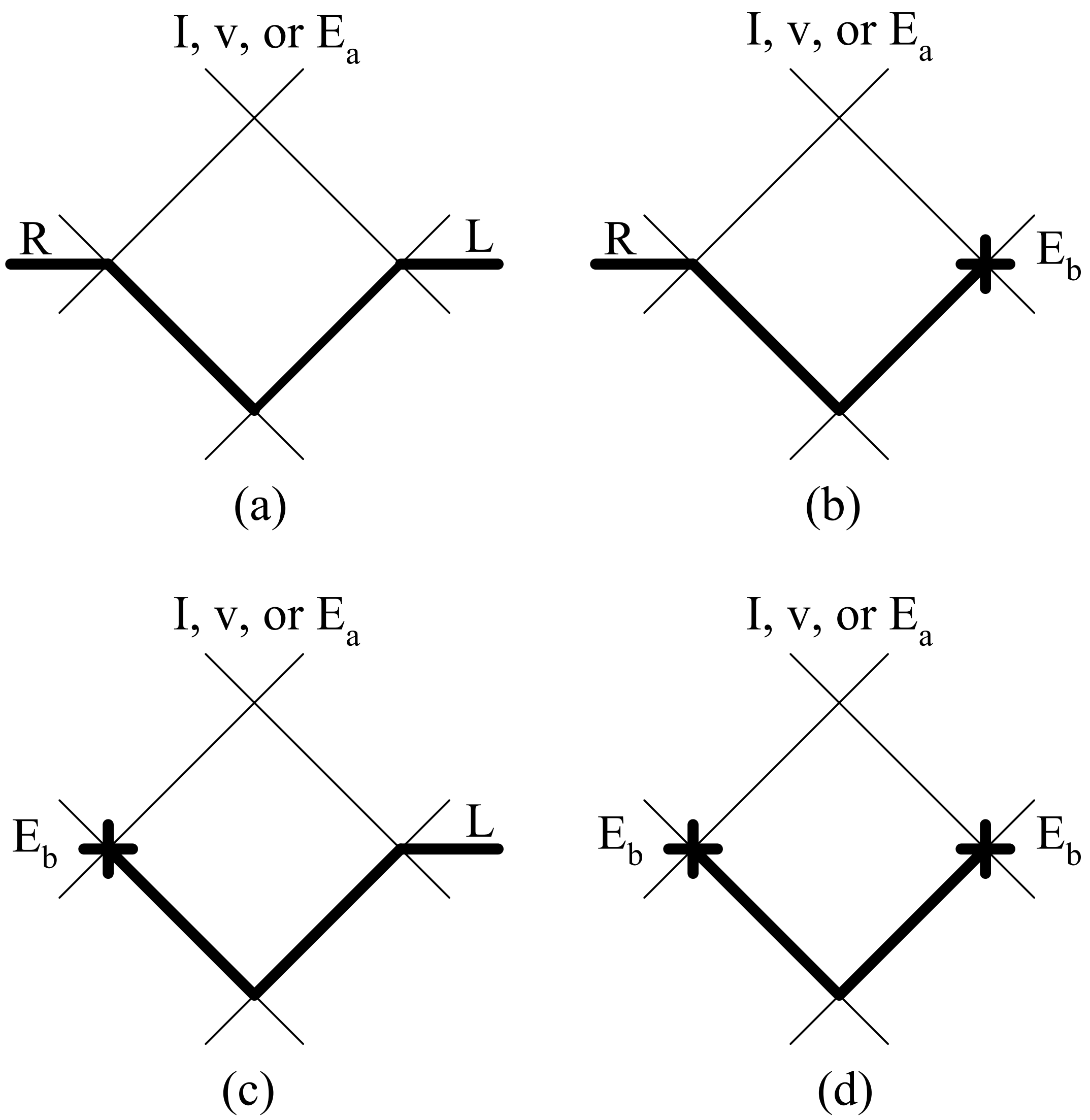

Figure 4 (cont'd)

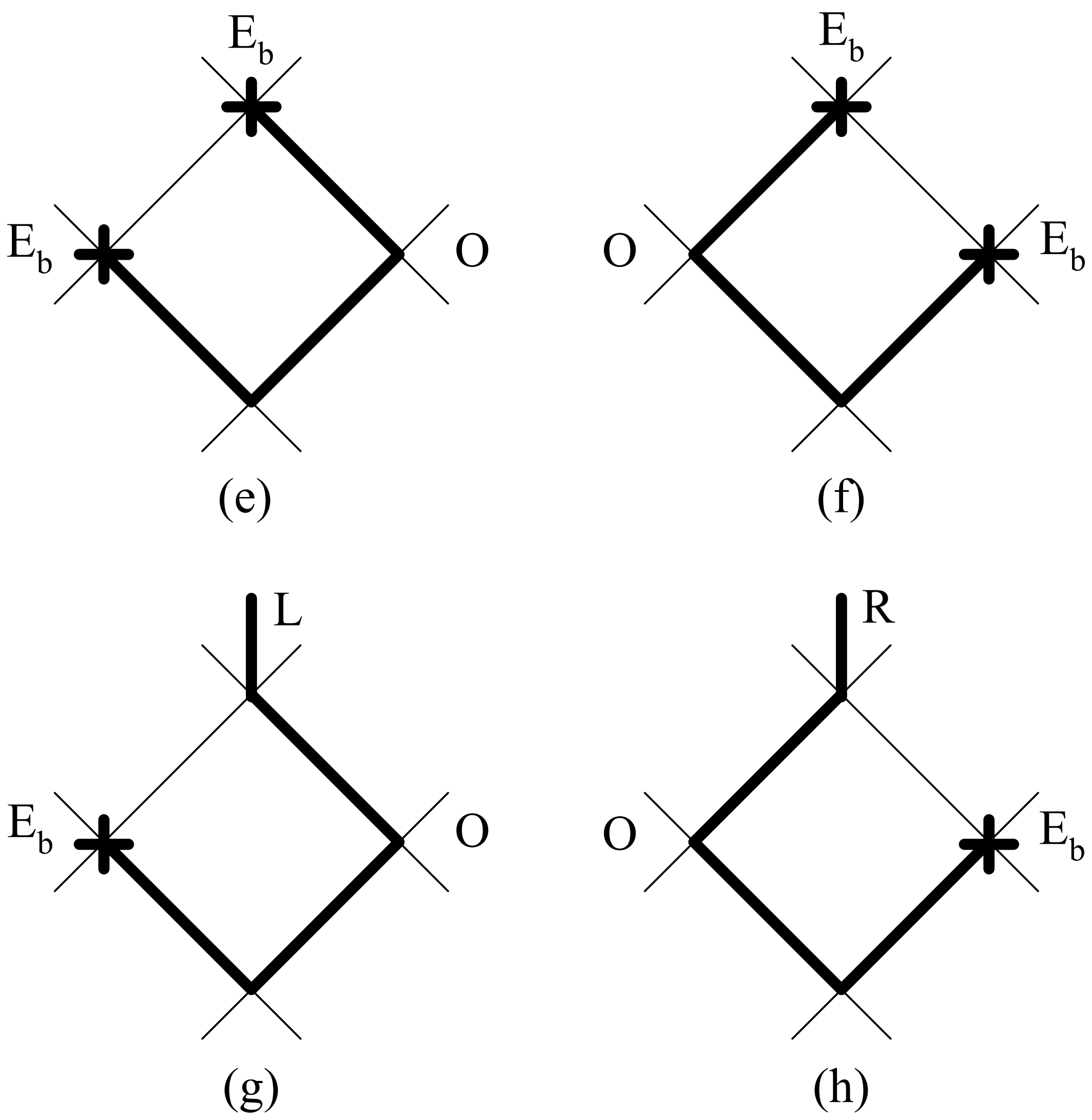

Figure 4 (cont'd)

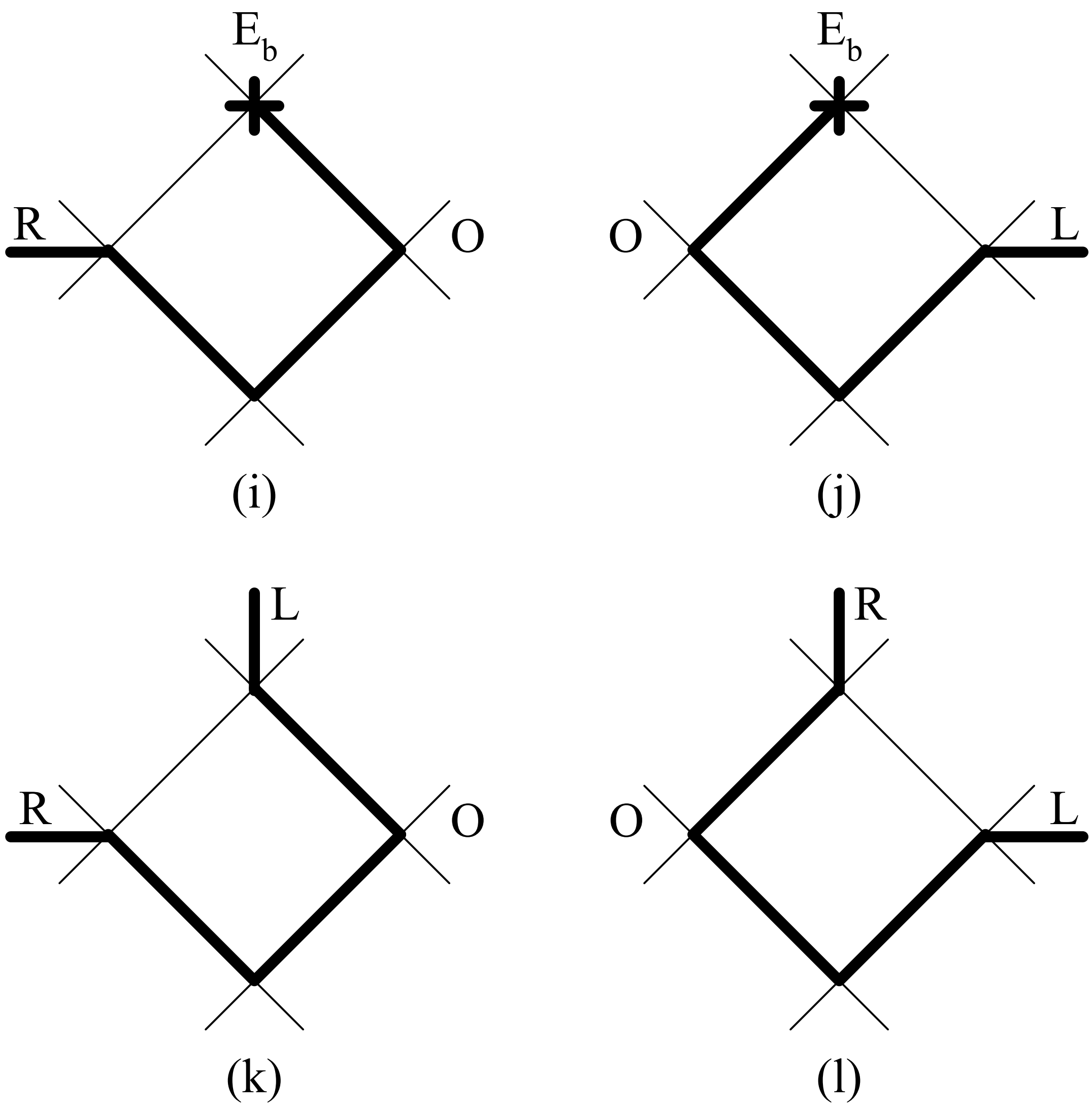

Figure 4

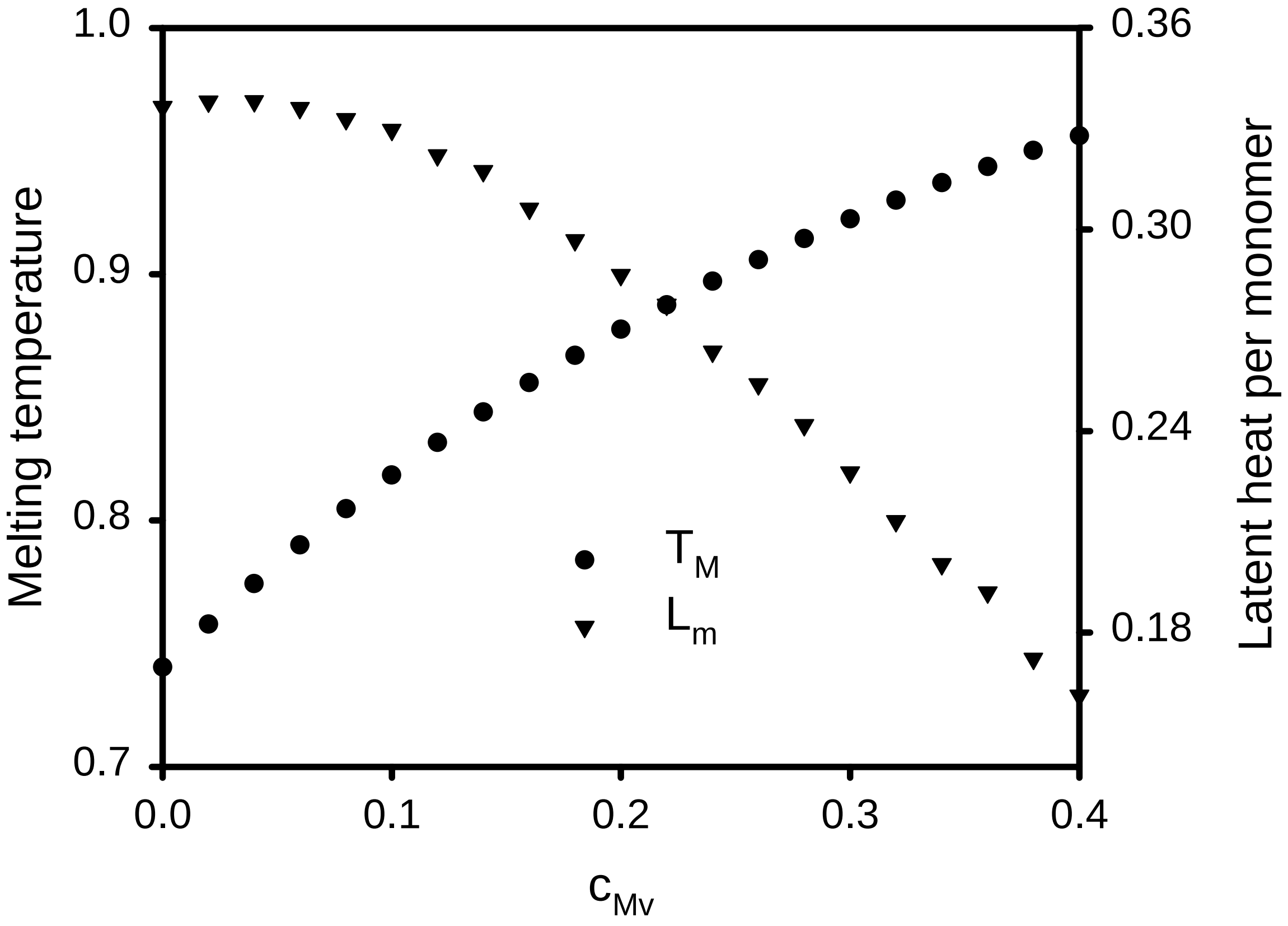

Figure 5(a)

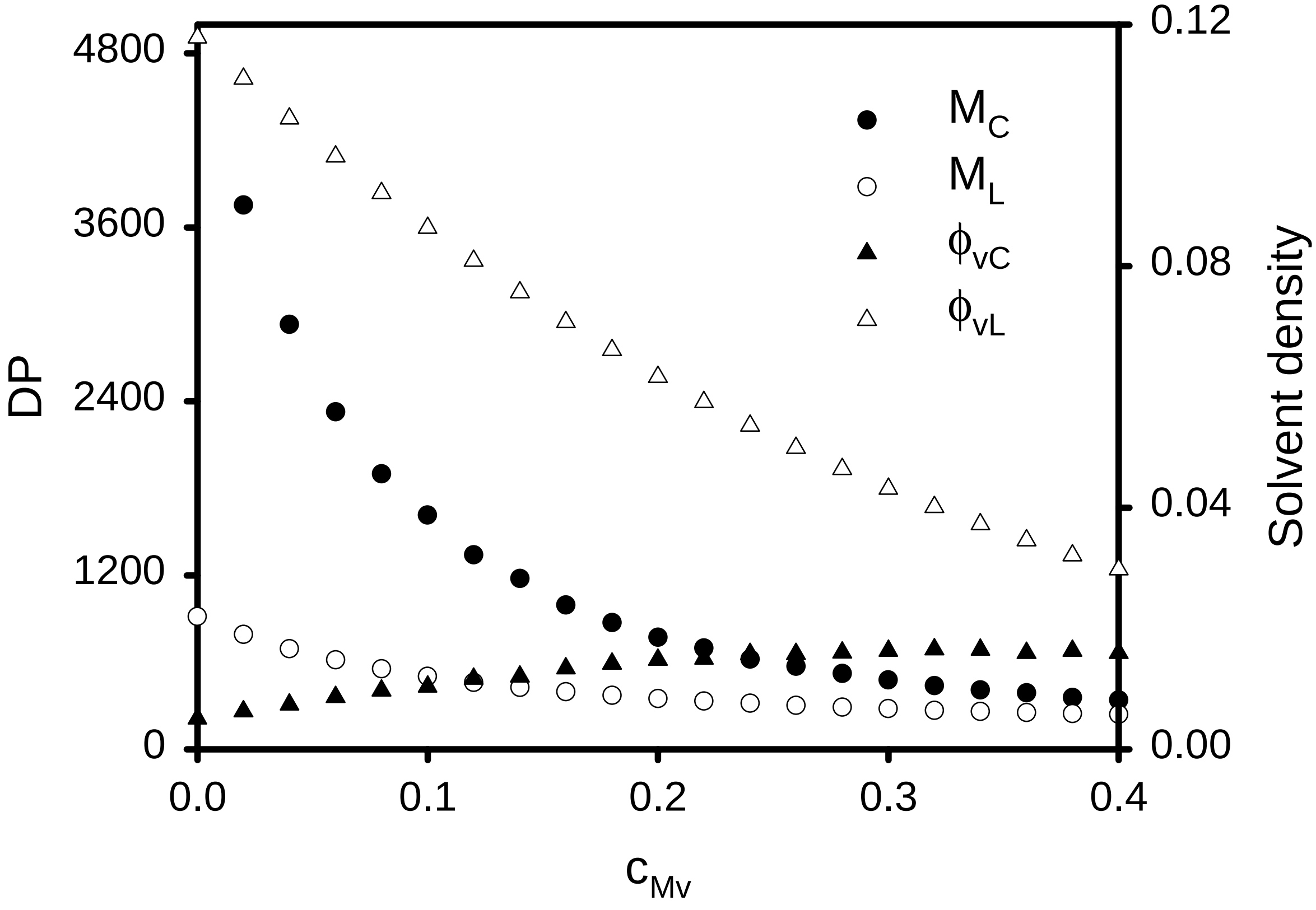

Figure 5(b)

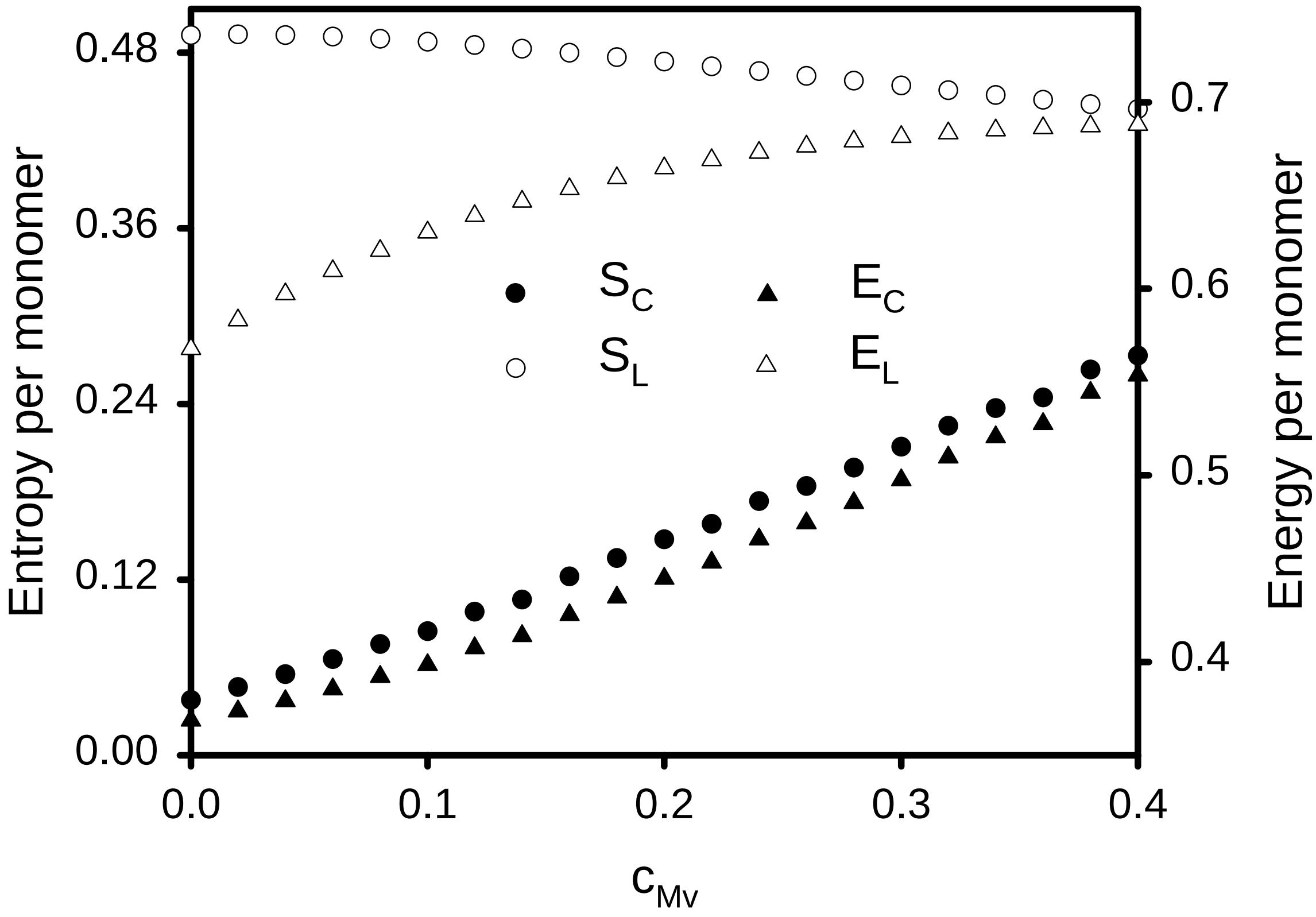

Figure 5(c)

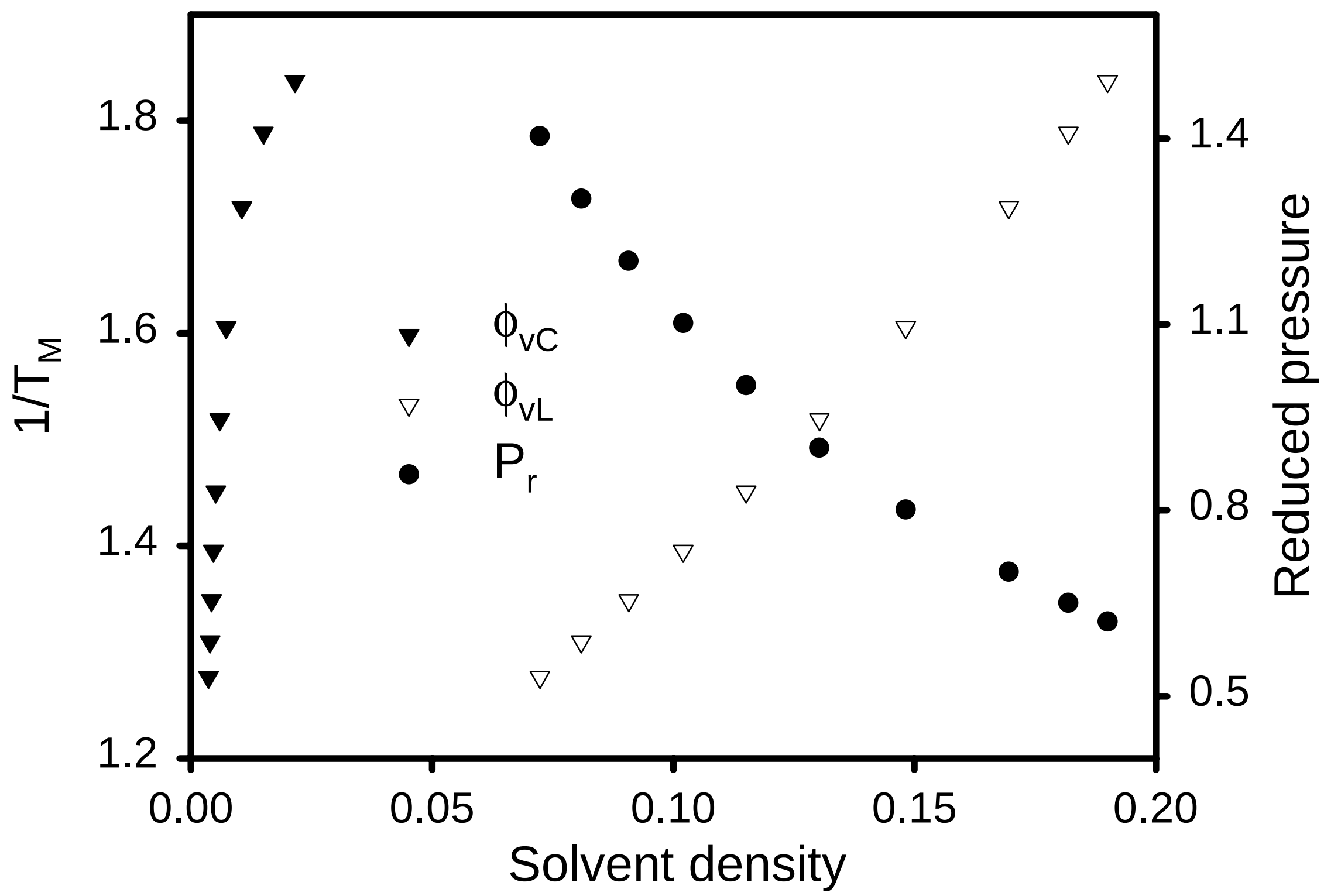

Figure 6(a)

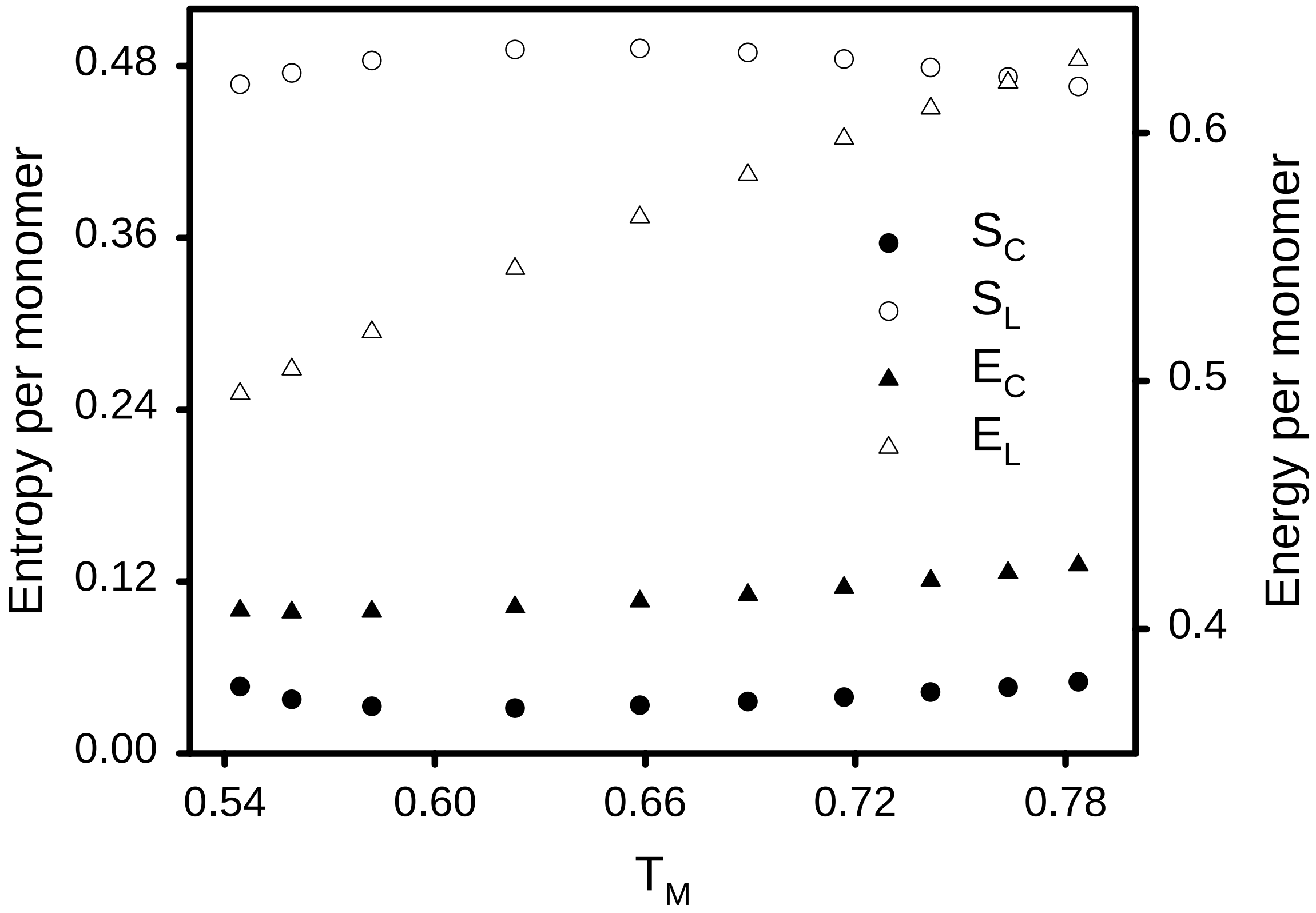

Figure 6(b)

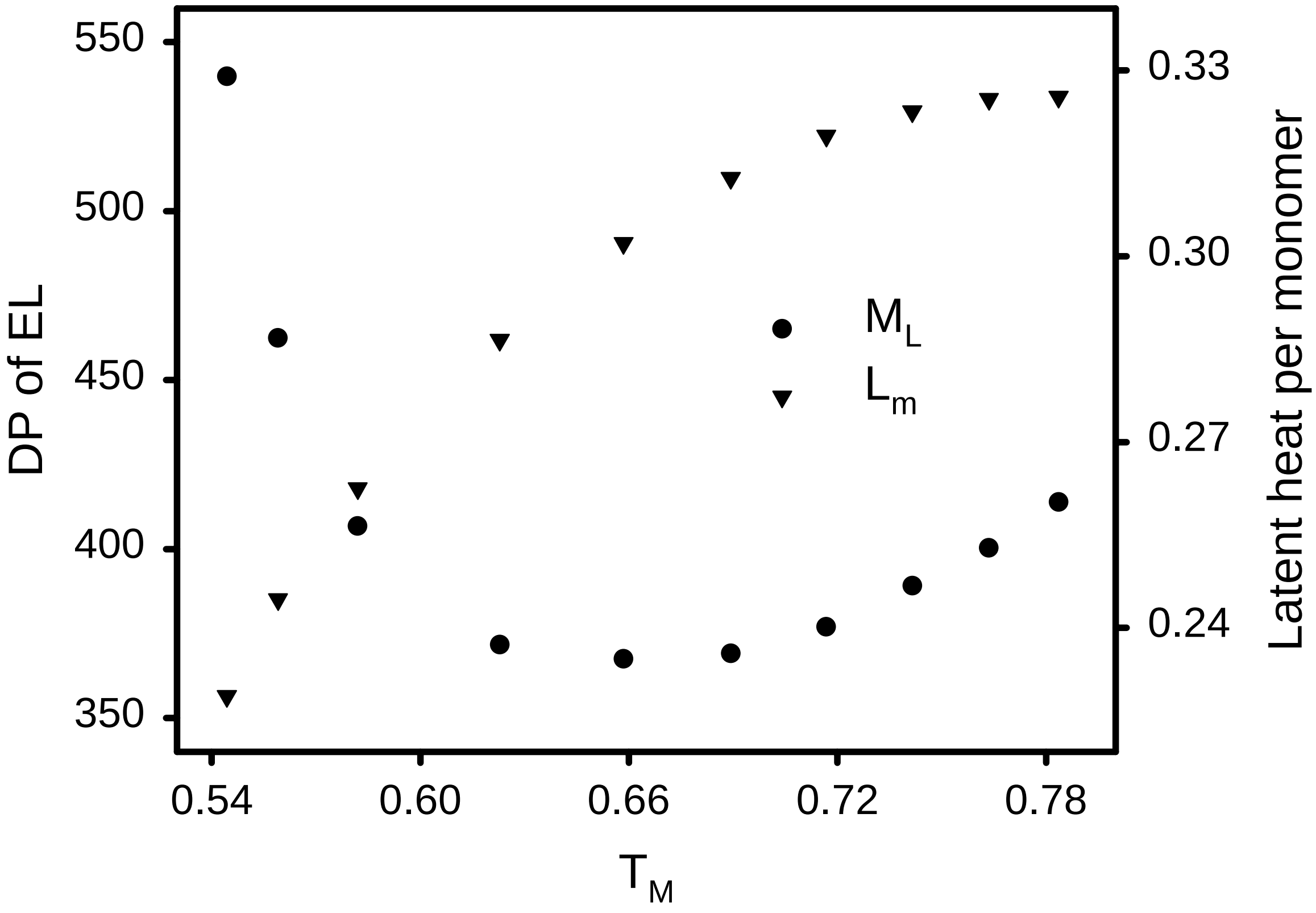

Figure 6(c)

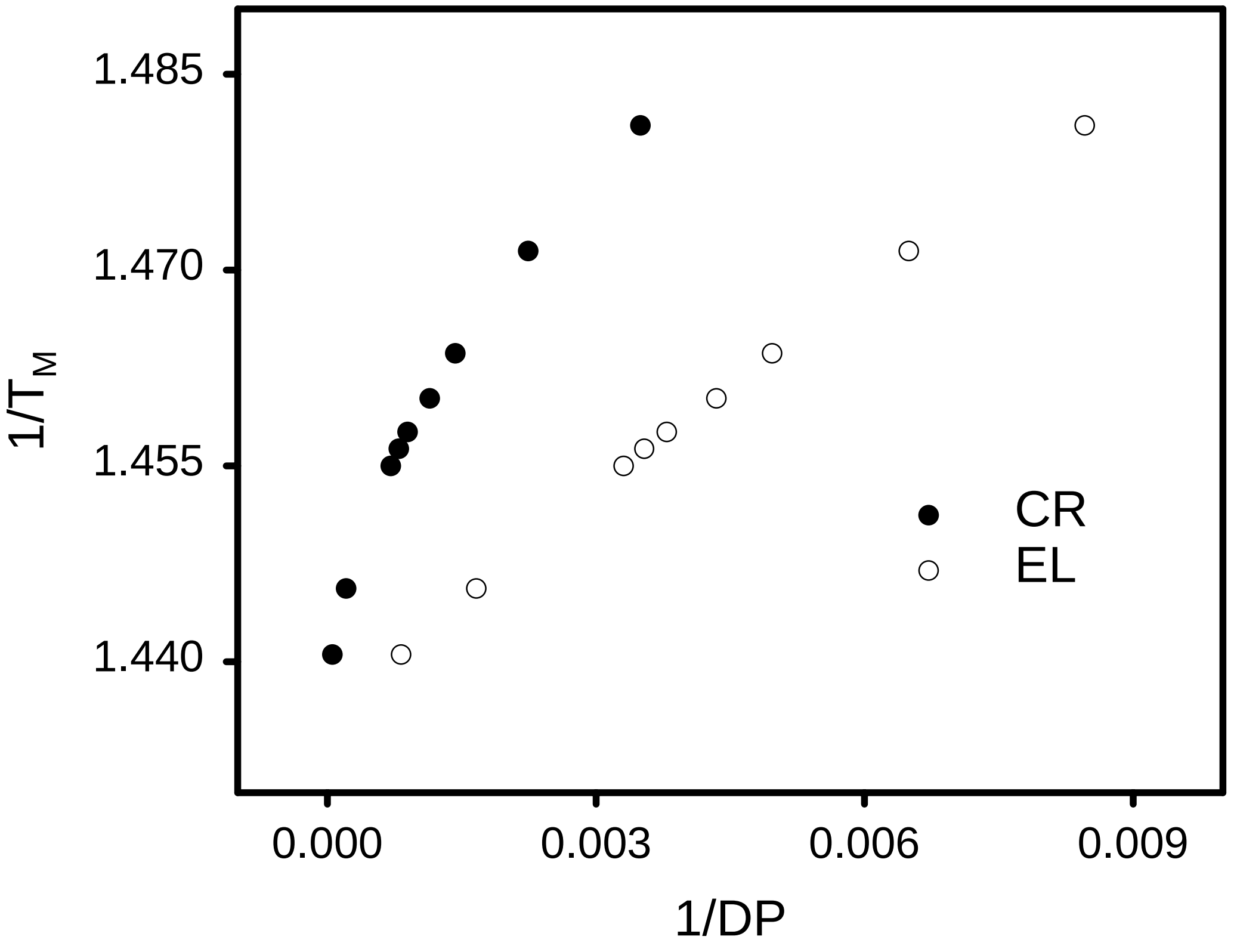

Figure 7(a)

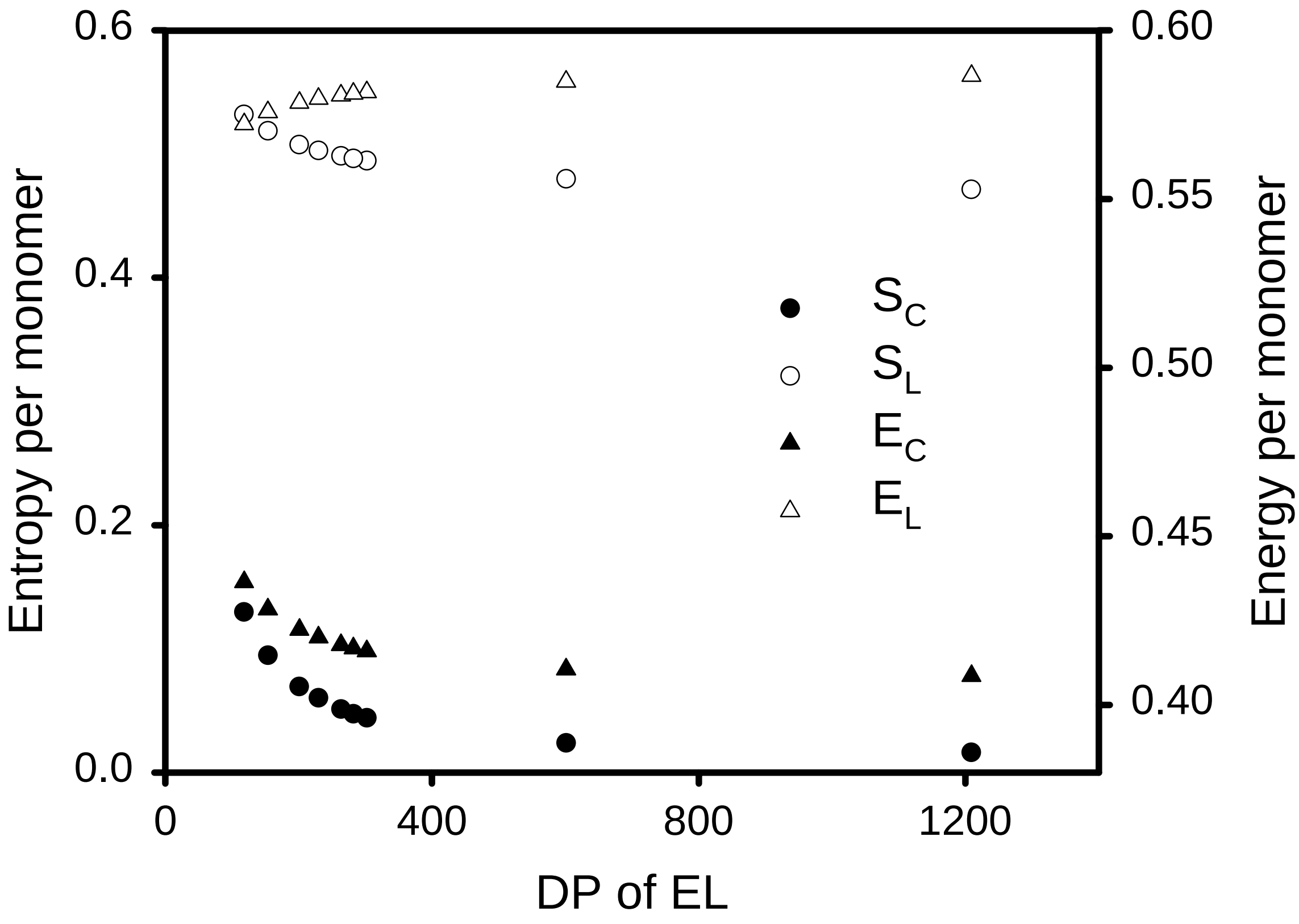

Figure 7(b)

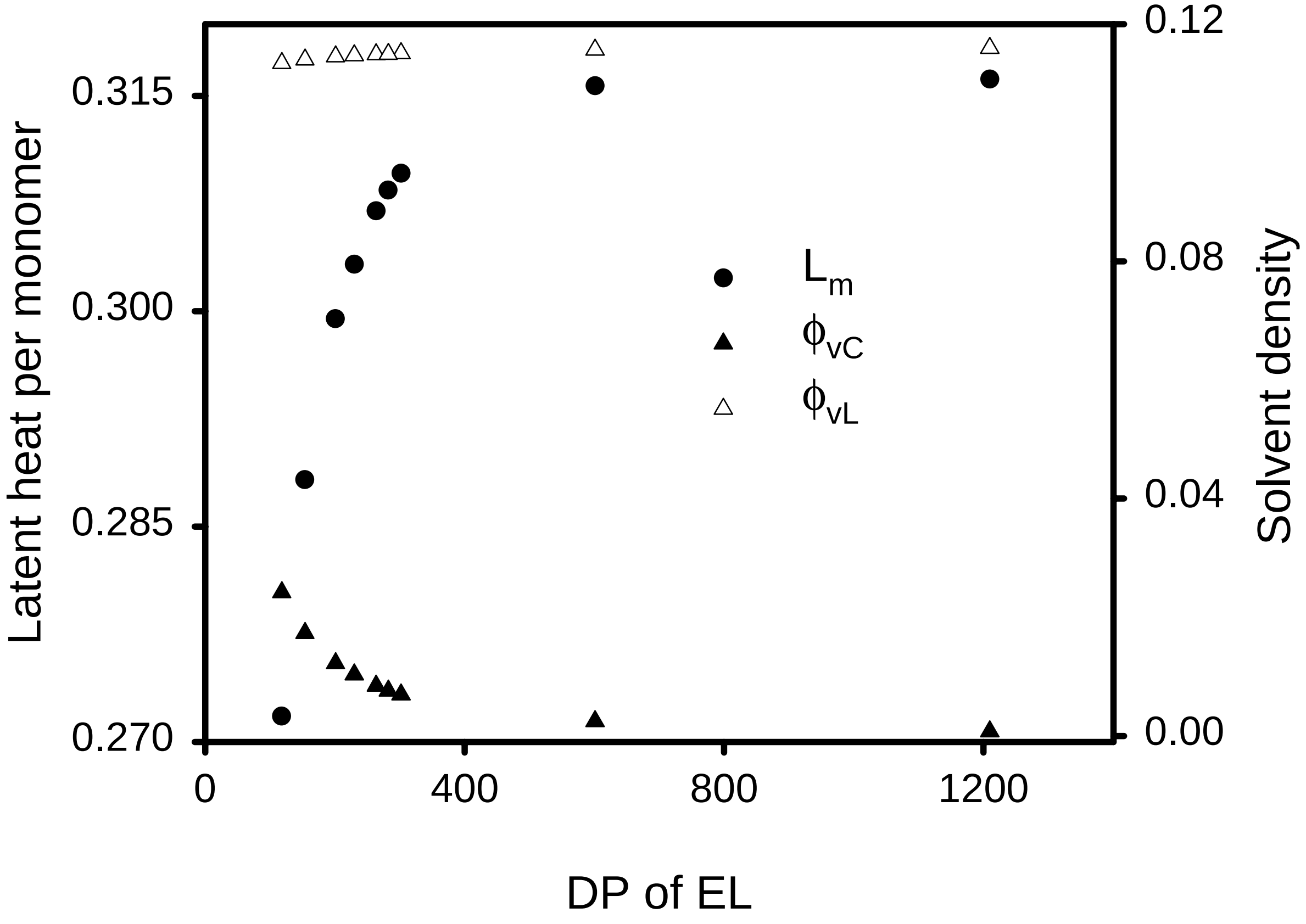

Figure 7(c)

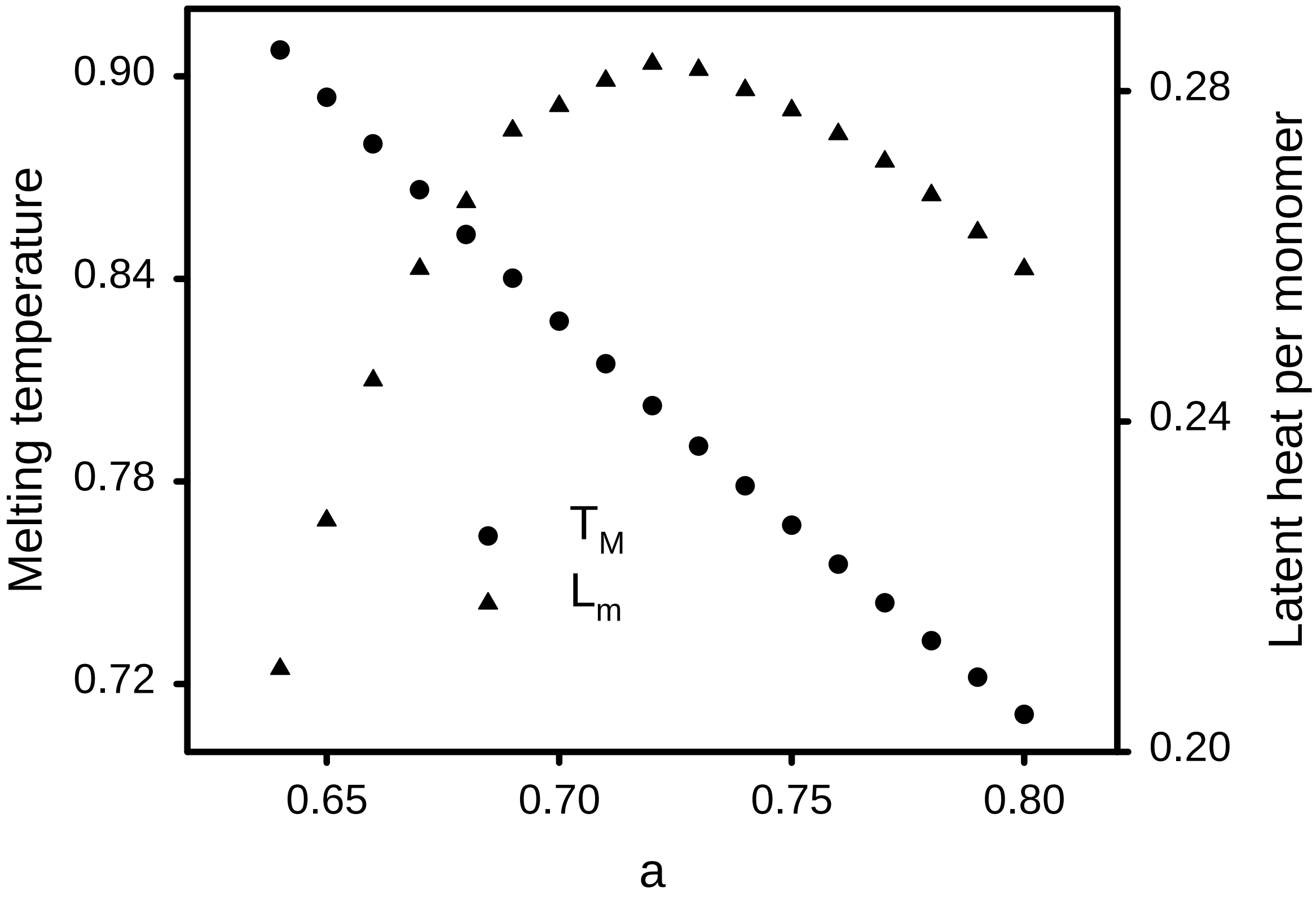

Figure 8(a)

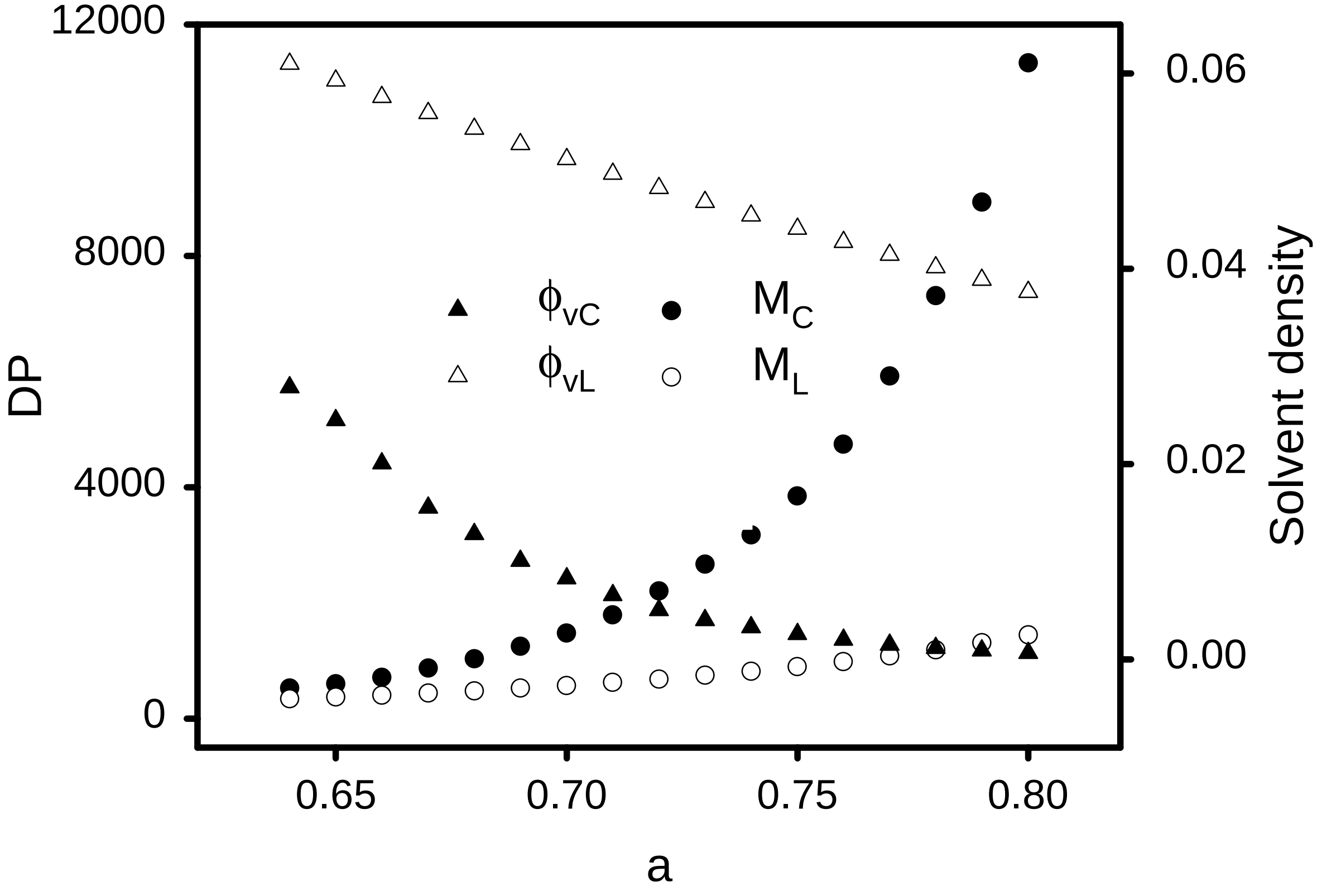

Figure 8(b)

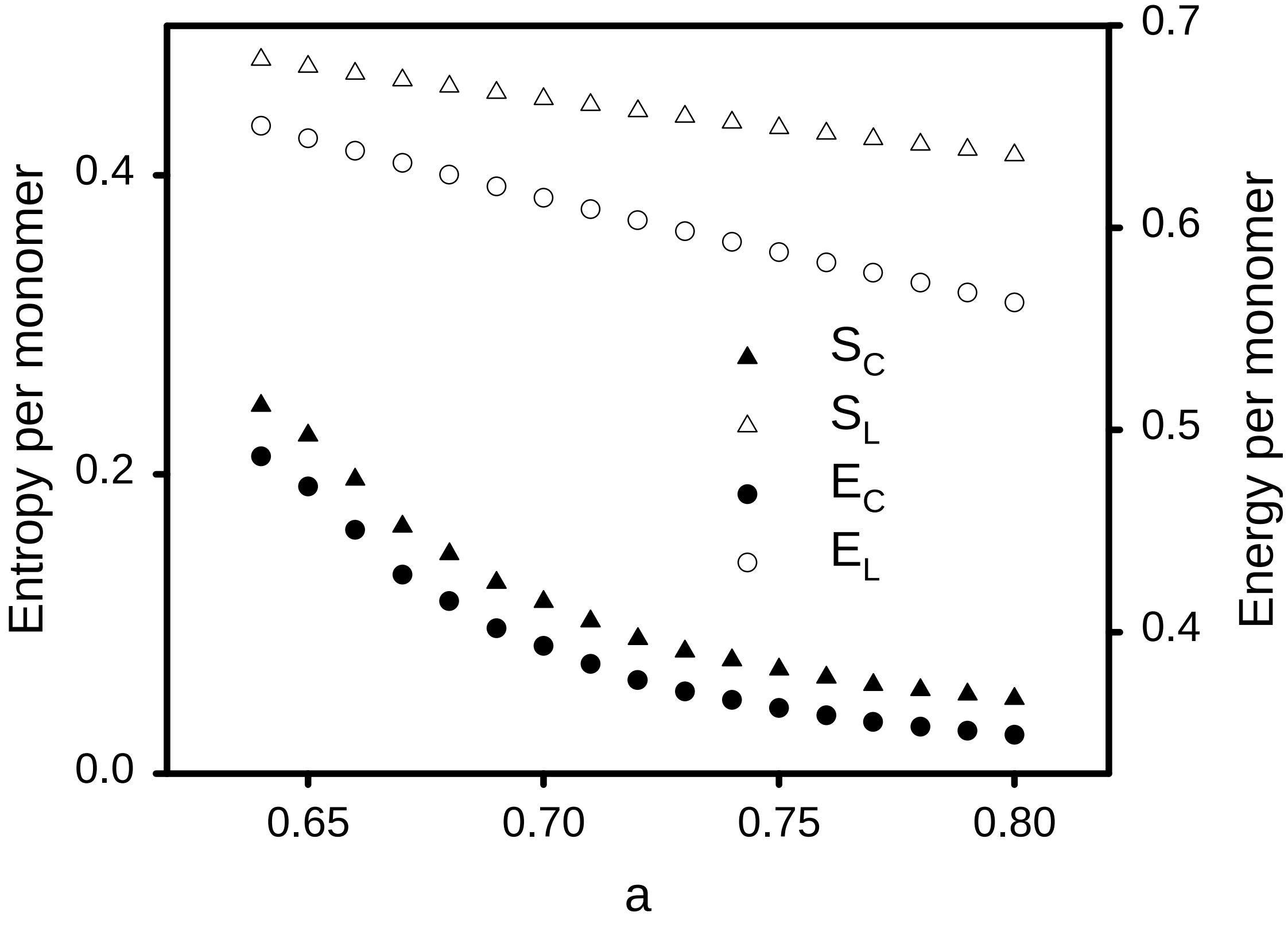

Figure 8(c)

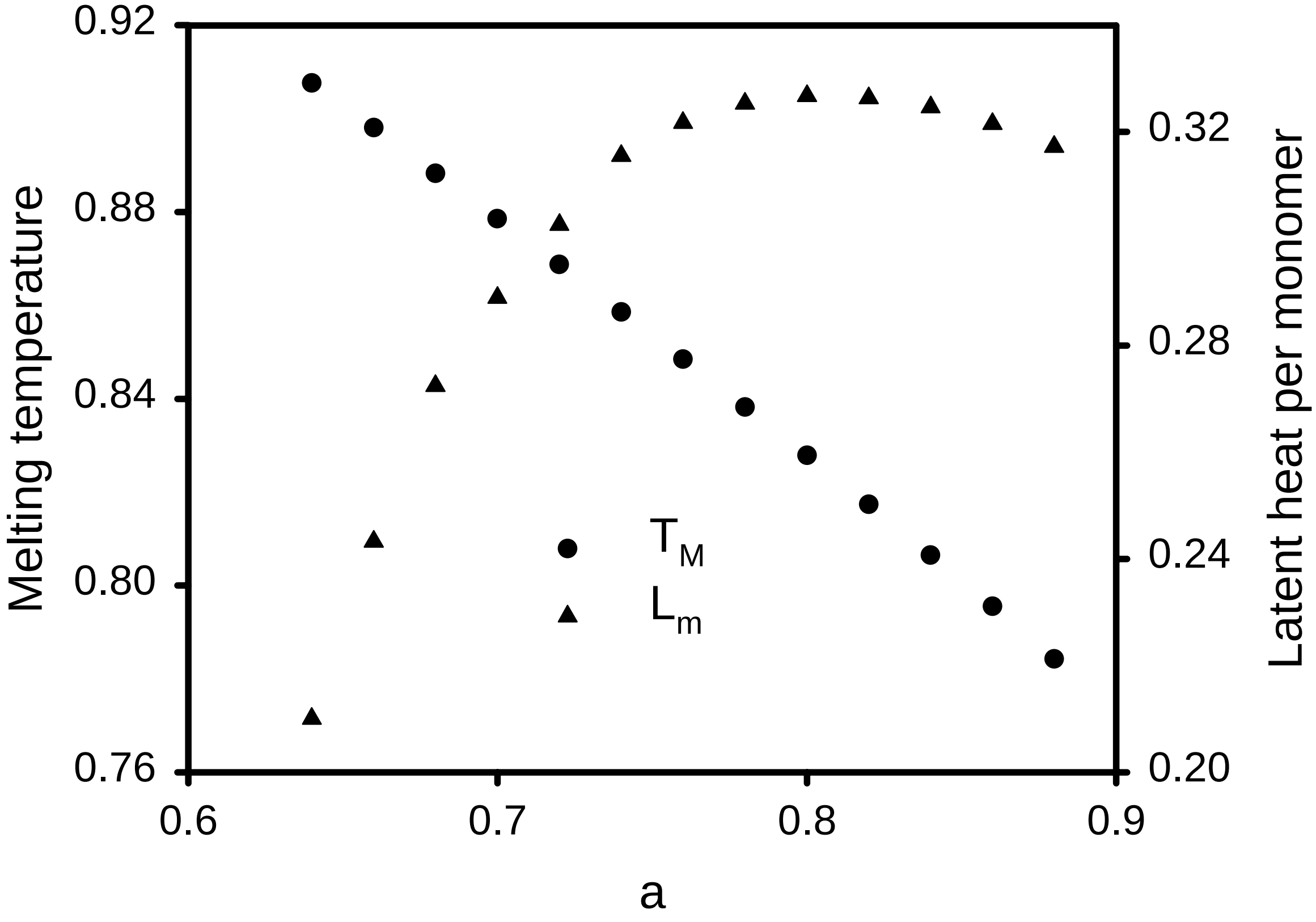

Figure 9(a)

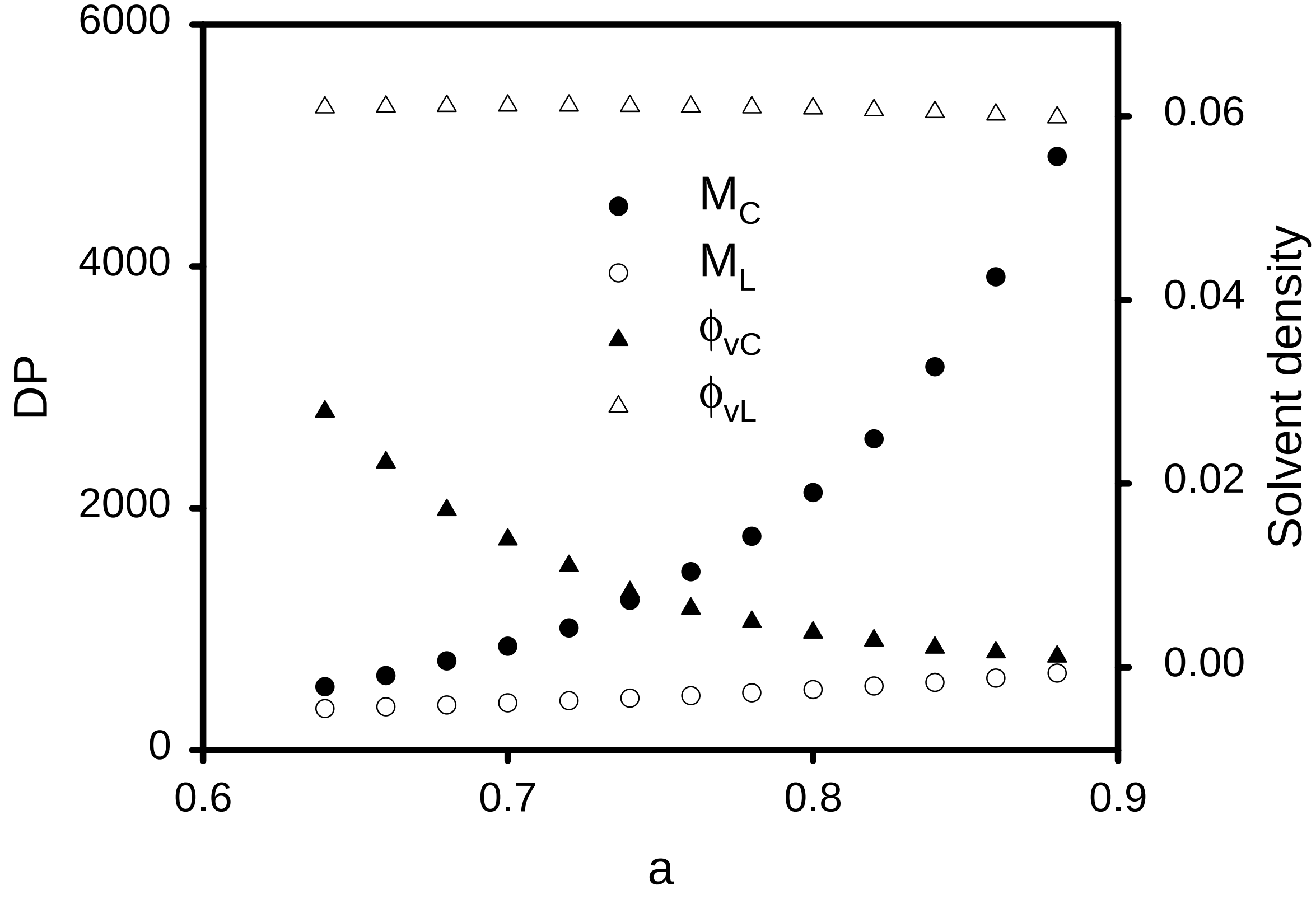

Figure 9(b)

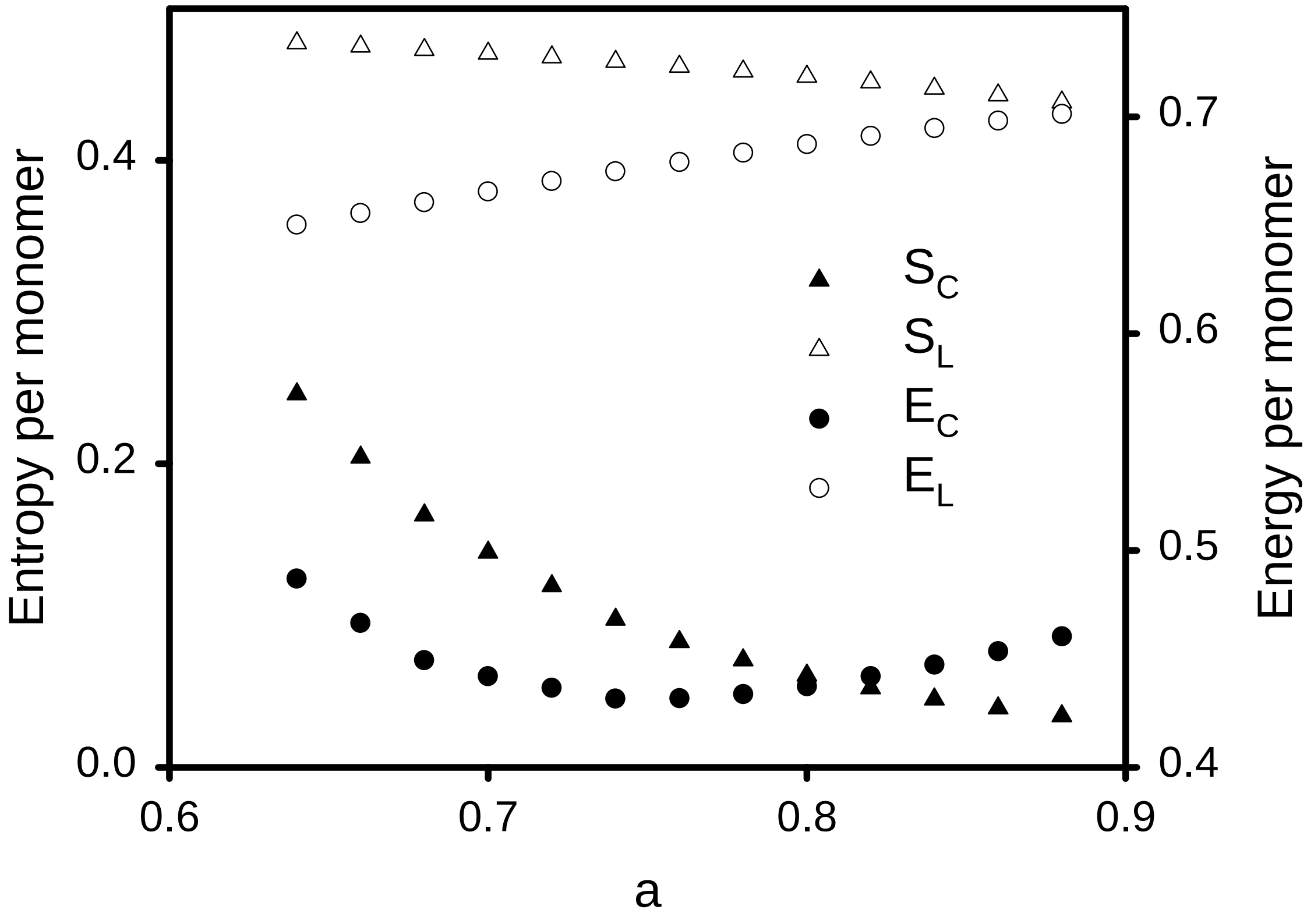

Figure 9(c)

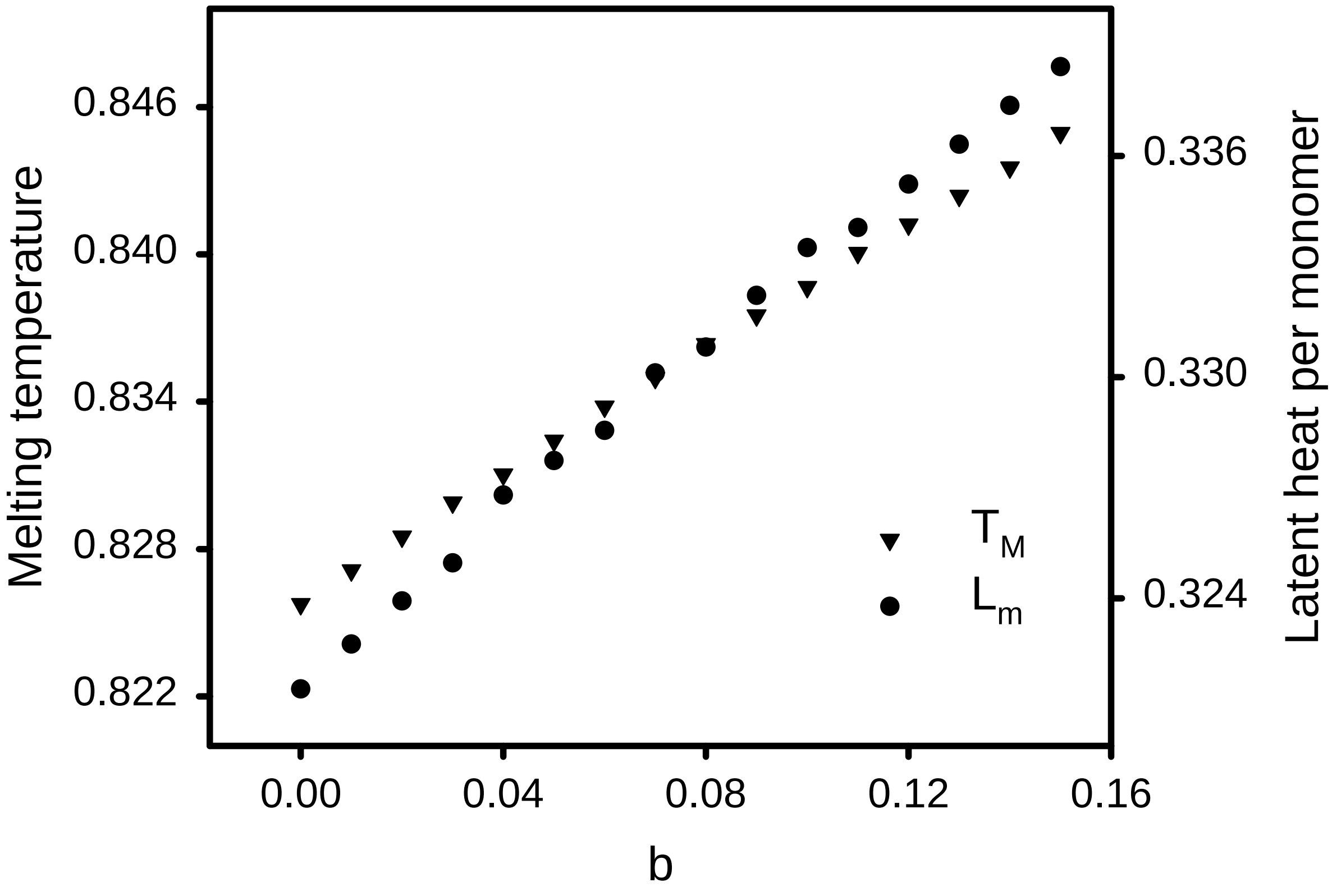

Figure 10(a)

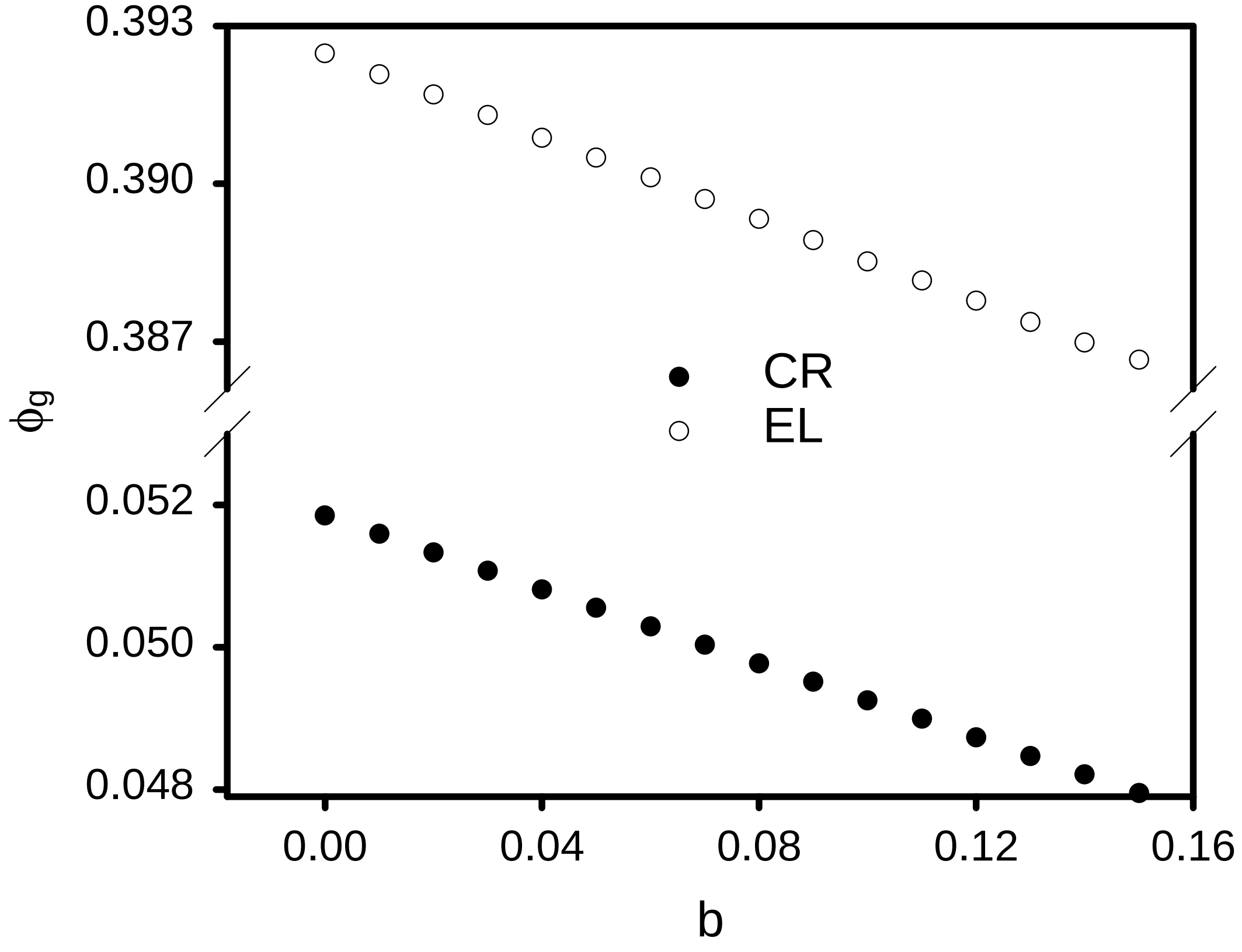

Figure 10(b)